\documentclass[preprint,showpacs,superscriptaddress,longbibliography,11pt]{revtex4-1}
\usepackage{diagbox}
\usepackage{amsmath}
\usepackage{amsfonts}
\usepackage{physics}
\usepackage[dvipsnames]{xcolor}
\usepackage[space]{grffile}
\usepackage{natbib}
\usepackage{multibib}
\usepackage{braket}
\usepackage{float}
\usepackage{color}
\usepackage{blindtext}
\usepackage{comment}
\usepackage[normalem]{ulem}
\newcommand{\STO}{SrTiO$_3$}

\newcommand{\bx}{{\mathbf x}}

\begin{document}

\title{Cavity-engineered ferroelectricity in \STO}

\author{Omar Mehio}
\thanks{These authors contributed equally to this work}
\affiliation{Kavli Institute at Cornell for Nanoscale Science, Cornell University, Ithaca, NY, USA}
\affiliation{School of Applied and Engineering Physics, Cornell University, Ithaca, NY, USA}
\affiliation{Laboratory of Atomic and Solid State Physics, Cornell University, Ithaca, NY, USA}

\author{Sylvia Griffitt}
\thanks{These authors contributed equally to this work}
\affiliation{Department of Physics, Cornell University, Ithaca, NY, USA}

\author{Andrew T. Pierce}
\affiliation{Kavli Institute at Cornell for Nanoscale Science, Cornell University, Ithaca, NY, USA}
\affiliation{Laboratory of Atomic and Solid State Physics, Cornell University, Ithaca, NY, USA}
\affiliation{Max Planck Institute for the Structure and Dynamics of Matter, Hamburg, Germany}

\author{Emil {Vi\~{n}as Bostr\"{o}m}}
\affiliation{Max Planck Institute for the Structure and Dynamics of Matter, Hamburg, Germany}

\author{Hanbyul Kim}
\affiliation{Department of Physics and Photon Science, Gwangju Institute of Science and Technology, Gwangju, Republic of Korea}

\author{Dylan P. Sotir}
\affiliation{Department of Materials Science and Engineering, Cornell University, Ithaca, NY, USA}

\author{Darrell G. Schlom}
\affiliation{Kavli Institute at Cornell for Nanoscale Science, Cornell University, Ithaca, NY, USA}
\affiliation{Department of Materials Science and Engineering, Cornell University, Ithaca, NY, USA}
\affiliation{Leibniz-Institut für Kristallzüchtung, Berlin, Germany}

\author{Simone Latini}
\affiliation{Max Planck Institute for the Structure and Dynamics of Matter, Hamburg, Germany}
\affiliation{Department of Physics, Technical University of Denmark, Kgs Lyngby, Denmark}

\author{Kin Fai Mak}
\affiliation{Kavli Institute at Cornell for Nanoscale Science, Cornell University, Ithaca, NY, USA}
\affiliation{Laboratory of Atomic and Solid State Physics, Cornell University, Ithaca, NY, USA}
\affiliation{Max Planck Institute for the Structure and Dynamics of Matter, Hamburg, Germany}
\affiliation{School of Applied and Engineering Physics, Cornell University, Ithaca, NY, USA}

\author{Jie Shan}
\affiliation{Kavli Institute at Cornell for Nanoscale Science, Cornell University, Ithaca, NY, USA}
\affiliation{School of Applied and Engineering Physics, Cornell University, Ithaca, NY, USA}
\affiliation{Max Planck Institute for the Structure and Dynamics of Matter, Hamburg, Germany}
\affiliation{Laboratory of Atomic and Solid State Physics, Cornell University, Ithaca, NY, USA}

\author{Dongbin Shin}
\affiliation{Max Planck Institute for the Structure and Dynamics of Matter, Hamburg, Germany}
\affiliation{Department of Physics and Photon Science, Gwangju Institute of Science and Technology, Gwangju, Republic of Korea}

\author{Angel Rubio}
\email[]{angel.rubio@mpsd.mpg.de}
\affiliation{Max Planck Institute for the Structure and Dynamics of Matter, Hamburg, Germany}
\affiliation{Initiative for Computational Catalysis (ICC), The Flatiron Institute, 162 Fifth Avenue, New York, NY 10010, USA}
\affiliation{Nano-Bio Spectroscopy Group, Universidad del País Vasco UPV/EHU, San Sebastián, Spain}

\author{Ankit Disa}
\email[]{asd47@cornell.edu}
\affiliation{Kavli Institute at Cornell for Nanoscale Science, Cornell University, Ithaca, NY, USA}
\affiliation{School of Applied and Engineering Physics, Cornell University, Ithaca, NY, USA}

\begingroup
\renewcommand{\baselinestretch}{1.2}
\maketitle
\endgroup

\noindent\textbf{Abstract}

\noindent Interfacing quantum materials with photonic cavity architectures is at the vanguard of materials engineering. The structured electromagnetic fluctuations of a dark cavity environment can renormalize fundamental thermodynamical material properties and energy scales, holding promise to create a rich ecosystem of cavity-induced ground states and functionalities absent in the bare material \cite{anappara_signatures_2009,niemczyk_circuit_2010,zhang_collective_2016,li_vacuum_2018,wang_evidence_2026}. However, direct experimental demonstrations that cavity coupling can stabilize or enhance genuinely new phases of matter have remained scarce. In this work, we establish the presence of cavity-induced and cavity-enhanced ferroelectric phases in free-standing \STO\ membranes coupled to dark THz-frequency photonic crystal cavities. Using THz spectroscopy, we observe ultrastrong coupling (USC) between the ferroelectric soft mode in \STO\ and the cavity resonance. The structural response of \STO\ to this USC is measured via second harmonic generation imaging, which reveals a substantial (greater than 400\%) enhancement of signal in the cavity environment and a distinct modification of the crystalline symmetry, indicating the onset and enhancement of ferroelectricity \cite{nova_metastable_2019}. We interpret these observations as resulting from renormalized piezoelectric and flexoelectric couplings, which in turn arise from the cavity-modified dielectric environment. This interpretation is supported with microscopic modeling via DFT-based multi-scale dynamics simulations and by macroscopic Ginzburg-Landau theory. Altogether, our experimental and theoretical results firmly confirm that cavity engineering can indeed be used to create new ground states in materials and enhance existing ones. This work establishes a new paradigm for materials engineering with direct applications for quantum transduction, quantum sensing, and nonlinear integrated photonics. 

\newpage
\noindent\textbf{Main Text}

\noindent\textit{Introduction}

Identifying novel pathways to engineer quantum materials is crucial to leverage their properties for device applications. For decades, light has proven to be a promising avenue towards achieving this control, as various light-matter interactions can be utilized to precisely tune microscopic material properties. These efforts have spanned both the electromagnetic spectrum and different timescales, yielding control over structural, electronic, and magnetic properties via mechanisms such as Floquet engineering, photodoping, and nonlinear phononics \cite{mikhaylovskiy_ultrafast_2015,wang_observation_2013,wall_ultrafast_2012,schoenlein_femtosecond_1987,juraschek_ultrafast_2017,fausti_light-induced_2011,li_keldysh_2022,forst_nonlinear_2011}. Each of these approaches leverage the light from classical light sources – real electric fields that interact with the material to drive a steady-state or nonequilibrium excited state of the material that may host different properties than the undriven system. 

A novel approach instead relies on the quantum nature of light – namely the structured vacuum electromagnetic fluctuations of a photonic cavity. The foundation of this approach is the idea that material properties are not determined by the material alone, but rather its interactions with the electromagnetic density of states that surround it \cite{ciuti_quantum_2005,wang_phase_1973,de_liberato_light-matter_2014}. This idea has long been verified in atomic cavity quantum electrodynamic (QED) systems, where fundamental atomic properties such as the spontaneous emission rate can be tuned via coupling to the photonic modes of the cavity \cite{sauer_cavity_2004,brennecke_cavity_2007,cao_qubit_2011}. In the solid state, these ideas can be taken much further, potentially yielding control over functional material properties and the collective phenomena of ordered ground states.

In a materials setting, it is possible to reach extreme coupling strengths that approach or exceed the fundamental energy scales of the system \cite{frisk_kockum_ultrastrong_2019}. In this so-called ultrastrong coupling (USC) regime, the light-matter interaction strength is no longer a perturbative effect but an active participant that can steer the ground state of the material as it competes with other microscopic interactions \cite{anappara_signatures_2009,niemczyk_circuit_2010,zhang_collective_2016,li_vacuum_2018}. Indeed, theoretical investigations have suggested that extreme interactions between material excitations and the structured vacuum fluctuations of a cavity can yield a rich ecosystem of cavity-mediated ground states absent in the bare material, including superconducting, magnetic, quantum Hall, superradiant, and ferroelectric phases \cite{lu_cavity-enhanced_2024, vinas_bostrom_controlling_2023,wang_cavity_2019,nataf_rashba_2019,Latini2021}. 

This approach is fundamentally different from the laser-based approaches that have dominated the optical control of materials. In the case of cavity-based engineering, the light and matter form a hybrid quantum system, from which novel functionalities emerge at the macroscopic level. This modality of control lends itself well to integration within quantum sensing, computing, and transduction hardware and nonlinear integrated photonic platforms. Moreover, from a technical perspective, the cavity-based approach bypasses many of the limitations of traditional laser-based approaches. As these changes are predicted to occur in cavities that are not illuminated by an external drive---so-called dark cavities---they can be manufactured as integrated structures and deployed in settings without optical access and without the heating created by an electromagnetic drive.   

Experimentally, the possibility of inducing or enhancing long-range order in cavity environments has only recently begun to be explored. Pioneering experiments have shown that material properties can indeed be altered in a cavity environment, but have predominantly reported suppression or modification of existing orders, including superconductivity, quantum Hall states, and correlated electronic states \cite{keren_cavity-altered_2026,graziotto_cavity_2026,appugliese_breakdown_2022,jarc_cavity-mediated_2023,Fausti2026,Gao2026,Junichiro2026}. More recently, however, experiments have provided evidence for cavity-enhanced superconductivity in NbSe$_{2}$ using split-ring resonators \cite{wang_evidence_2026, Zhang_NbSe2_2026}, as well as an enhanced superconducting response in an underdoped cuprate in a tunable terahertz cavity \cite{Montanaro_2026}. These results establish cavity coupling as a viable means of enhancing existing ordered states, while the broader question of whether cavity interactions can induce genuinely new ground states remains open. Identifying platforms in which cavity interactions yield the formation of new phases of matter, or qualitatively enhance existing ones, is therefore crucial for establishing cavity engineering as a new modality of materials control. 

In this work, we establish the presence of cavity-induced and cavity-enhanced ferroelectric phases in ultra high-quality free-standing \STO\ membranes coupled to dark THz-frequency photonic crystal cavities. Using THz spectroscopy, we observe ultrastrong coupling (USC) between the ferroelectric soft mode in \STO\ and the cavity resonance, as well as a significant renormalization of the dielectric function. By coupling free-standing membranes of \STO\ to diamond-based one-dimensional THz photonic crystal cavities, we clearly observe the cavity-mediated formation and enhancement of ferroelectric states. These states are identified via second harmonic generation (SHG), which is a direct off-resonant probe of the structural response of \STO\ embedded in the cavity. By studying the dielectric response with THz spectroscopy, we determine that these changes are driven by USC between the cavity modes and the ferroelectric soft mode (FES), which in turn modifies the piezoelectric and flexoelectric couplings that determine the ferroelectric domain structure of \STO. The renormalization of these coupling constants is further supported by microscopic DFT-based multi-scale dynamics simulations and macroscopic Ginzburg-Landau calculations. \\ 

\newpage
\noindent\textit{Cavity-\STO\ Experimental Platform}

\STO\ is a paradigmatic quantum paraelectric (QPE)---a system in close proximity to a ferroelectric critical point, but in which long-range ferroelectric order is frustrated by zero-point quantum fluctuations of the FES (Figure 1a)~\cite{Shin2021}. In the bulk, \STO\ remains centrosymmetric and paraelectric at all temperatures. Signatures of the QPE state emerge at low temperatures, where the dielectric constant and the FES frequency exhibit a dramatic increase and softening, respectively, before saturating below $\sim$ 35 K \cite{muller_srtimathrmo_3_1979}.

This strongly fluctuating system is the ideal testbed for cavity materials engineering, as a THz-frequency cavity mode can couple directly to the dipole-active FES, which ranges in frequency from 2.5 THz at room temperature to 0.5 THz at low temperatures. In the USC regime, dressing of the FES by the cavity field can renormalize its effective mass and suppress zero-point fluctuations \cite{Latini2021}. Interaction with the cavity thus pushes the system towards the ferroelectric instability (Figure 1b). These effects are enhanced by the proximity to the ferroelectric transition, as the FES is endowed with a divergent oscillator strength ($d$) in this critical region, increasing the light-matter coupling strength ($g$) by several orders of magnitude as compared to typical platforms for cavity QED, where $g/\omega = d/\sqrt{2\hbar\varepsilon_{0}\omega V}$, with cavity frequency $\omega$ and mode volume $V$. These strongly enhanced interactions raise the prospect of significant mode renormalization through vacuum fluctuations alone, without any external drive. Indeed, theoretical calculations predict that coupling to a dark cavity can induce a QPE-to-ferroelectric phase transition via both non-resonant dynamical localization and resonant phonon-polariton softening \cite{Latini2021,ashida_quantum_2020}.

However, while the QPE properties of \STO\ make it a promising candidate for cavity engineering, they also create several technical challenges that need to be overcome when considering the device architecture. The dramatic enhancement of the dielectric constant, exceeding 10,000 in bulk samples, blocks the transmission of light at low frequencies via Fresnel reflection, creating a significant technical hurdle for the coupling of high-$Q$ THz cavity devices to \STO\ crystals \cite{yang_epitaxial_2022}. Moreover, the QPE state is known to be fragile to strain, defects, and external stimuli, setting stringent limits on the material quality, fabrication, and integration within the device \cite{muller_srtimathrmo_3_1979,sirenko_soft-mode_2000}. Finally, to probe the effect of the cavity on the FES and the ferroelectric properties requires in situ spectroscopic access to the cavity-embedded \STO\ in both the THz and visible frequency range. Combining this sample access with the need for sub-wavelength mode confinement to enhance the light-matter coupling strength adds strong constraints to the cavity components and architecture.  

We solve these issues through a rational co-design approach, balancing the technical requirements of the material, photonic, and spectroscopic subcomponents of the device architecture. The resulting device architecture is shown in Figure 1c. The cavity platform is a one-dimensional photonic crystal cavity (PCC) composed of alternating layers of diamond and air. Diamond is chosen because of its weak absorption and flat dispersion at THz frequencies, its transparency to optical probes, and because its temperature-independent index of refraction ($n\sim 2.4$ at 1 THz) and minimal thermal expansion ensure stability of the cavity mode with temperature \cite{floch_electromagnetic_2011,xiao_terahertz_2023,stoupin_thermal_2010}. The PCC consists of a $\lambda/4$ THz Bragg reflector with a defect at its center, created by doubling the thickness of the central diamond plate. The fundamental mode of the cavity has a peak field amplitude occurring at the surfaces of the defect layer, the frequency of which can be tuned by changing the thicknesses of the air and diamond layers. Figure 1d shows the spectroscopic response of the cavity, showcasing its temperature independence and large ($>10$) $Q$-factor. 

Introducing bulk \STO\ into the PCC would obstruct the cavity mode due to its large THz reflectivity. To circumvent this problem, yet still enable strong cavity coupling, we consider the finite penetration of light within its reflectance band. In Extended Data Figure 1, the simulated transmission spectra of a $\sim$1.5 THz cavity---resonant with the FES at 100 K---with different thicknesses of \STO\ placed on the defect layer are shown (Methods). While thicker \STO\ layers completely suppress the cavity mode, below roughly 100 nm polaritonic coupling between the cavity mode and the FES emerges, indicating that the cavity mode is well-preserved despite the introduction of the material. Hence, coupling a thin ($<$ 100 nm) film of \STO\ into our device provides a pathway to explore cavity-induced effects. Experimentally, however, growing high-quality nanoscale \STO\ films onto diamond is a challenge due to the large epitaxial mismatch. 

To solve this problem, we fabricated free-standing membranes of \STO\ using oxide molecular beam epitaxy (Methods). Following Refs. \cite{zhang_super-tetragonal_2024,chiu_presence_2022,lu_synthesis_2016}, free-standing membranes are fabricated by introducing a selectively etchable buffer layer (Sr$_{4}$Al$_{2}$O$_{7}$) between the growth substrate and the target \STO\ thin film during growth (Extended Data Figure 2). After etching the buffer layer in its solvent (water), the \STO\ film is released and transferred directly onto the diamond PCC defect layer using the standard polymer dry-stamping techniques used routinely to assemble heterostructures of 2D materials (Methods).

Characterization of a $\sim$37 nm-thick \STO\ membrane via time-domain THz spectroscopy (TDTS) reveals a strong absorption that we identify as the FES (Figures 1e,f). The FES strongly disperses with temperature, softening from its room temperature value of around 2.5 THz towards a minimum value of $\sim$0.5 THz around 60 K. Below 60 K, the mode begins to harden slightly indicating the formation of ferroelectricity. The linewidth also increases significantly, likely due to inhomogeneous broadening. Such inhomogeneity is expected for membranes, which are prone to strong strain gradients caused by stretching and wrinkling upon transfer that can lead to inhomogeneity in the ferroelectric polarization among different domains, and hence different values for the FES below the transition temperature \cite{li_classical--quantum_2025}. 

In order to directly visualize the local polarization structure in the \STO\ membranes, we use rotational-anisotropy second harmonic generation imaging (RA-SHG), which allows us to map the spatial distribution and symmetry of ferroelectric domains.  In Figures 1g and 1h, characteristic SHG images of the sample are shown for two orthogonal incident light polarizations. These images clearly indicate the formation of several ferroelectric domains, aligned along different crystal axes, as confirmed by the two-lobe RA-SHG patterns shown in Figure 1i. In addition, we identify regions with distinct four-lobe RA-SHG patterns that are weaker than in the ferroelectric regions (Figure 1j). We identify these as regions with strongly multi-domain character, reminiscent of the polar nanoregions of the QPE phase. Comparison of the temperature dependencies in the ferroelectric and these QPE-like regions further confirms this picture, as the former demonstrates a clear order-parameter-like onset of signal that is much weaker in the latter (Figure 1k). 

From these results, we can draw the conclusion that the \STO\ membranes have mixed ferroelectric and QPE or multi-domain character. Such a device provides an ideal platform for studying cavity QED in this setting, as it enables us to investigate, in one and the same sample, how cavity coupling affects ferroelectricity, quantum paraelectricity, and domain structure. A second important conclusion is that TDTS and RA-SHG together define a holistic characterization suite that allow us to characterize the both the dielectric and structural properties of the \STO\ both in and out of the cavity. These probes will allow us to quantitatively determine whether cavity coupling can indeed create or enhance ferroelectric order in \STO.\\

\noindent\textit{Dielectric Response of Cavity-\STO}

Having characterized the bare properties of the photonic crystal cavities and the \STO\ membranes, we now consider the dynamics of the coupled system. We start by considering the dielectric response to directly visualize the interaction of the cavity mode with the FES. In Figure 2a, we show the THz transmission of a cavity centered at 1.5 THz coupled to a $\sim$37 nm membrane of \STO, measured with TDTS as a function of temperature (Methods). The bare response of the particular membrane used in this device is shown Extended Data Figure 3, and the bare response of the cavity is shown in Figure 1d. The dotted lines are linear fits through the uncoupled dispersions of the bare FES and cavity resonance. In the coupled system, clear avoided crossing behavior emerges, indicating strong coupling between the cavity mode and the FES. The excitations that form the two branches in the coupled spectrum are thus the upper phonon polariton (UP) and lower polariton (LP). 

The splitting between the UP and LP at the avoided crossing is equal to $2g$. The line cut at the avoided crossing---142 K---is plotted in Figure 2b, indicated that the UP and LP are separated by $\sim$0.7 THz. Given that the cavity and FES frequencies are 1.5 THz in the uncoupled system at this temperature, the normalized light-matter coupling strength is $\sim$0.2, well into the USC regime. These results were reproduced in a separate cavity device, as shown in Extended Data Figure 4. 

To better understand the FES-polariton, we use the bare measured dielectric constant of the \STO\ membrane in a transfer matrix simulation of the cavity response (Methods, Extended Data Figure 3). We construct a multilayer model accounting for the diamond, air, and \STO\ layers and calculate the absolute transmission in the same geometry as the experiment. As shown in Figure 2c, a polaritonic splitting also emerges in the simulation. The experimentally observed spectra are overlaid over the simulation as white curves, demonstrating very strong agreement between the two. 

At low temperatures, near the ferroelectric transition, the FES and 1.5 THz cavity mode are still strongly hybridized, obscuring any changes to the STO dielectric function. To understand how cavity coupling affects the FES near the FE instability, we fabricated a separate cavity device with a central frequency of 2.6 THz. Since the cavity mode is off-resonant with the FES at low temperatures, we can investigate the dispersion of the FES in a cavity environment without the complication of polariton formation in the frequency regime relevant to the ferroelectric transition. The bare temperature-dependent transmission spectra of the \STO\ membrane fabricated for this device demonstrates a FES that softens to 0.5 THz and broadens significantly below roughly 70 K, as expected for an inhomogeneously broadened membrane (Figure 3a). Upon coupling to the cavity, the FES demonstrates a marked change (Figure 3b). Not only does it sharpen significantly, but its dispersion is different, hardening below 70 K. As before, we can use the experimental dielectric function of the bare membrane (Extended Data Figure 5) to simulate the expected response (Figure 3c, Methods). The simulated response again captures the major features of the cavity coupled system, demonstrating the same hardening and sharpening behaviors that were measured in the experimental spectra. 

The mode sharpening demonstrated in the off-resonant cavity is strongly suggestive that modification of the ferroelectric phase is possible in our \STO-PCC platform. A prerequisite for the cavity to modify ferroelectricity is that it couples to and modifies the dielectric properties of the material, as this phase is characterized by a spontaneous polarization. In the present system, the USC between the FES and the cavity resonance not only modifies the dispersion in the case where the FES and the cavity mode are resonant---forming a phonon polariton---but also when the two modes are strongly off-resonant. 
However, confirming that the ferroelectric properties are indeed modified by this coupling requires an \textit{in situ} measurement of the structural properties of the \STO\ membrane. \\

\noindent\textit{Structural Response of Cavity-\STO}

Having established the impacts of cavity coupling on the dielectric properties of \STO, we now turn towards characterizing its effects on the ferroelectric order using RA-SHG imaging. Due to the modular design of the PCC, we are able to repeatably characterize the response of the same regions of the sample in both the bare and cavity-coupled geometry, allowing for quantitative comparisons of both SHG intensity and polarimetric symmetry (Methods, Extended Data Figure 6). In Figures 4a and 4b, we show the bare and cavity-coupled SHG response of a 32 nm \STO\ membrane, respectively. For this device, the bare cavity resonance is centered at 1.5 THz and the coupled device demonstrates USC with $g/\omega$ exceeding 0.1 (Extended Data Figure 4). A stark contrast can be seen between the bare and coupled response of the membrane, with the clearest difference being a dramatic increase of SHG intensity in the cavity setting, exceeding 400\% in certain regions. In the particular region shown in Figure 4, the enhancement is rather uniform, affecting the majority of the image as shown in the intensity difference map in Extended Data Figure 7b. However, in other regions of the sample, we also see areas of decreasing intensity or mixed character (Extended Data Figure 7). These differences in intensity are ubiquitously observed over multiple devices with both 1.5 THz and 2.5 THz central frequencies (Extended Data Figure 7). 

In addition to these changes in SHG intensity, the cavity-coupled system also demonstrates regions with modified crystalline symmetry as observed through changes in the RA-SHG pattern. In Figures 4d and 4f, the bare and coupled response of a spot that is QPE in the bare sample is shown. Whereas the spot features four nearly equal lobes in the bare sample, two dominant lobes form in the coupled device. This shift towards a two-fold symmetric pattern, coupled with the large increase in SHG intensity, indicates that coupling to the cavity alters the crystalline symmetry towards a strongly polarized ferroelectric state. On the other hand, other regions show no change in crystalline symmetry despite a large enhancement or suppression in SHG symmetry (Figures 4c, 4e). We note that neither of these changes is associated with a significant change in the ferroelectric-to-paraelectric critical temperature within our resolution of 5 K (Figure 4g). 

To better understand the distribution of these symmetry modifications and their relationship to the intensity changes, we perform a spatial mapping of the symmetry in the bare and coupled samples (Methods). This mapping is performed by binning the image into $0.7\ \mu\text{m}\  \times\ 0.7\ \mu\text{m}$ regions and fitting the polar plot obtained from each region with a phenomenological model of the RA-SHG response \cite{denev_probing_2011,nova_metastable_2019}: 
\begin{equation}
I_{\rm SHG}(\phi) \propto \left|A_1\cos^{2}(\phi) + A_2\sin^{2}(\phi) + B\cos(\phi)\sin(\phi)\right|^{2}
\end{equation}
where $\phi$ is the polar angle formed by rotating the polarization of the probe pulse with respect to the crystal axes and $A_1$, $A_2$, and $B$ are fitting constants (Methods). An important feature of this model is that it creates a clear criterion to distinguish between different domains by considering the relative values of $A_1$ and $A_2$. To do this, we define the following symmetry parameter: 
\begin{equation}
S = |A_1| - |A_2|.
\end{equation}
to classify domains within each region and create a spatial domain map (Methods). When $S$ is positive (negative), it indicates that there is a ferroelectric domain with lobes along $0^{\circ}$ ($90^{\circ}$). On the other hand, when $S=0$, the pattern with four equal lobes emerges, indicating a highly fluctuating or paraelectric domain. 

Figure 5 displays the results of this spatial symmetry analysis. To best highlight the observed response, we chose a characteristic area with a clear domain boundary and regions where the SHG intensity is both enhanced and suppressed. However, the results are general and apply to all measured devices, including the devices presented in Figure 4 and Extended Data Figure 7. The bare and coupled responses of this region are shown in Figure 5a and 5b, respectively. In the bare response, two large dominant domains can be seen with opposite $S$, separated by a thin domain wall with $S=0$. To the left of the $S>0$ domain, an additional domain of largely $S=0$ character can be seen, with small polarized pockets. In the cavity, the shape of each domain is clearly modified, as highlighted by the $\Delta S$ map shown in Figure 5c obtained by taking the difference between the two maps. Importantly, many different behaviors can be observed in this map, with some regions increasing in polarization as $S$ is pushed away from 0 in the cavity, while others demonstrate depolarization behavior, with $S$ shifting towards 0 in the cavity. There is also a large region with very minimal $\Delta S$, exhibiting very weak striations with symmetry changes spatially oscillating in opposite directions. The SHG intensity difference $\Delta I$ between the cavity-coupled and the bare membrane is shown in Figure 5d. Again, a varied phenomenology is observed, with regions demonstrating sharp, oscillating striations of intensity increases and decreases, while others exhibit more uniform changes to the intensity.  

To draw connections between the spatial variations in both the symmetry and intensity channels, we highlight in Figures 5e–h characteristic SHG patterns from select spots that capture the overall observed phenomenology. The chosen spots are marked in Figures 5a-d. Starting with the region polarized by the cavity, the spot shown in Figure 5e is another example of a clear symmetry change from a four-lobe $S=0$ pattern towards a ferroelectric $S=1$ pattern. Like the region plotted in Figure 4d, this increase in polarization is also associated with a large increase in intensity, as expected for a transition into a ferroelectric state. On the other hand, in Figure 5f, we show a spot that demonstrates a transition from a two-lobe $S=1$ pattern to a four-fold $S=0$ pattern, indicating that the cavity has depolarized a region. From this, we can conclude that the cavity is capable of both inducing and suppressing ferroelectric order in the very same sample. 

The sharp striations in $\Delta S$ and $\Delta I$ occur in the same region of the sample, and the response in these regions is ferroelectric in both the bare and the cavity-coupled membrane. The oscillations have nearly the same period and shape in both maps, indicating that they are related. The spot shown in Figure 5g shows a marked increase in SHG intensity, but with a weak positive change in $S$, indicating an enhancement of a ferroelectric domain with no symmetry change. Adjacent to this spot is the region shown in Figure 5h, which demonstrates the opposite behavior: a large decrease in intensity accompanied by a very weak negative $\Delta S$. 

We hypothesize that the origin of this varied response of the ferroelectric is that the cavity acts differently in regions with dominant piezoelectric or flexoelectric couplings. The relationship between polarization $\textbf{P}$ and strain $\boldsymbol{\epsilon}$ is:
\begin{equation}
    P_i = d_{ijk} \epsilon_{jk} + f_{ijkl} \frac{\partial \epsilon_{jk}}{\partial x_l}
\end{equation}
where $\mathbf{d}$ is the piezoelectric tensor and $\mathbf{f}$ is the flexoelectric tensor. The motivation behind this conjecture was our observation that regions with uniform $S$ generally display a suppression of ferroelectric polarization when the sample is placed in the cavity (Extended Data Figure 7). The uniformity of $S$ in these regions implies a large strain that sets a dominant ferroelectric polarization domain, but a relatively weaker strain gradient. On the other hand, we generally observe that regions with either rapidly varying or vanishing $S$ display an enhancement in ferroelectric polarization (Extended Data Figure 7). Unlike the uniform regions, these rapidly varying or weakly polarized regions likely feature strain gradients (flexoelectric coupling) that dominate over the strain. Since piezoelectricity relates strain to polarization while flexoelectricity relates the strain gradient to the polarization, we hypothesize that the cavity suppresses ferroelectricity in regions dominated by the former and enhances ferroelectricity in regions dominated by the latter, explaining the differences between uniform and rapidly varying regions. 

To support this hypothesis, we used atomic force microscopy (AFM) to create a topographic map of the sample. Though AFM does not measure strain directly, wrinkles and cracks in these samples create strain fields that are reflected in the topographic maps. The topographic map for the region plotted in Figures 5a-d is shown in Figure 5i, in which several long cracks can be observed. A particular crack---highlighted in red in Figure 5i and marked by the dotted line in Figure 5d---exactly tracks the oscillatory features seen in the $\Delta S$ and $\Delta I$ difference maps. Since the oscillations seem to stem from this crack, we can infer that the crack is creating a non-uniform strain field in the sample that affects the cavity-induced response. Given the presence of the oscillations in the difference maps, we can assume that the strain field is also oscillatory in nature. An oscillatory strain field would result in the maximum of the strain occurring when the strain gradient is 0, and vice versa. Under these conditions, the hypothesis that the cavity field increases the ferroelectricity when the flexoelectric coupling is dominant, and decreases it when the piezoelectric coupling is dominant, yields the same pattern that we observe---a periodic pattern of intensity increases and decreases, associated with corresponding enhancements and suppressions of $S$. 

To further confirm this picture, we also generated a topographic AFM map of a region that demonstrates a uniformly decreasing intensity in the cavity and a uniform $S>0$ (Extended Data Figure 7). This region is completely flat in the AFM, revealing no major topographic features. From the uniform SHG response and the smooth topography, we can assume that this region is uniformly strained and has minimal flexoelectric contributions. Our observation that this region shows a decrease in SHG intensity in the cavity thus further confirms that the cavity suppresses polarization when the region is piezoelectric-dominated, and that a flexoelectric component is needed to observe an enhancement in polarization. \\

\noindent\textit{Microscopic modeling}

To better understand the experimental observations, we theoretically investigate the cavity-induced modification of ferroelectricity in SrTiO$_3$ using a multiscale simulation framework based on density functional theory (DFT).
The anharmonic potential energy surfaces for the $a$-axis ferroelectric soft mode ($Q_\mathrm{f}$) and the $a$-axis lattice constant displacement ($u$) were extracted from DFT total energy calculations to determine the microscopic interaction parameters. 
Based on these, we constructed a microscopic Hamiltonian for the lattice dynamics of SrTiO$_3$, including nonlinear and inter-site phonon interactions (see the Supplementary Information (SI) for details of the phonon-QED Hamiltonian including flexoelectricity). 
Quantum fluctuations of the ferroelectric soft mode and the cavity photons were described by a ring-polymer molecular dynamics (RPMD) scheme (details are provided in the SI).
We first simulated the real-time dynamics of a 1D chain of SrTiO$_3$ unit cells, with a prescribed strain profile used to mimic experimental conditions (upper panel of Fig.~6a).
For the first $50$~ps the light-matter coupling was taken to vanish, during which we found quantum paraelectric or weak ferroelectric behavior induced by a finite flexoelectric coupling. 
After 50~ps, once the coupling to the photon field (with $\omega_c = 1.5$~THz and $A_0 = 0.42$ a.u.) was switched on, domains with positive (red) and negative (blue) ferroelectric polarization were found to develop. This indicates that the coupling to the photon field localizes the wavefunction of the FES, and thereby enhances ferroelectricity, consistently with previous reports ~\cite{Latini2021} (see the SI for details of dynamical localization).
Additional polarization dynamics induced by a light-matter coupling, under various initial conditions, are presented in the SI.

To capture the larger-scale dynamics of the FES dynamics, we used the microscopic model to derive an effective continuum theory. Within this theory (Methods), the FES dynamics is described by a non-linear stochastic wave equation, accounting both for the internal dynamics and the coupling to a spatially dependent strain profile. With this mesoscopic framework, we simulated 2-D samples of STO reaching dimensions of $\sim 10 \times 10$ $\mu$m$^2$.

In Fig.~6, we simulate the effect of the cavity on a sample with constant strain regions connected by regions with finite strain gradients. The direct effect of the cavity is described as a suppression of the FES kinetic energy~\cite{Latini2021} (Methods). As the bare FES system is inversion symmetric, the mean polarization vanishes when statistically averaged, and the main effect of the cavity is a narrowing of the probability distribution $\mathcal{P}(Q)$ of the FES coordinate (SI, Figure S8). However, finite strain gradients break inversion symmetry by selecting a preferred polarization direction, and produce a nonzero mean polarization in regions with finite flexoelectric coupling (see Fig.~6b and SI Figure S11). In these regions, the cavity is found to increase the magnitude of ferroelectric polarization.

We also observed a reduction of ferroelectric polarization in regions of constant strain. Here, the local flexoelectric field vanishes, and the polarization is governed by a strain-dependent local potential together with spatial coupling to neighboring polarized regions. As the reduction also happens in regions between strain gradients that favor the same polarization direction, it cannot be explained solely by cancellation between neighboring regions with opposite flexoelectric biases.

The ferroelectric polarization can be connected to the measured SHG signal by writing the SHG intensity as $I_\mathrm{SHG} = I_\mathrm{QPE} + I_\mathrm{FE} \langle P\rangle^2$. Here $I_\mathrm{QPE}$ is a weak background contribution from sample imperfections, present in the QPE (or paraelectric) regime, while $I_\mathrm{FE}$ is the electric dipole contribution activated in the FE regime. Fig.~6c shows the difference in SHG intensity, $\Delta I_\mathrm{SHG} = I_\mathrm{SHG}(\text{cavity}) - I_\mathrm{SHG}(\text{no cavity})$, clearly displaying regions of both strong enhancement and suppression of ferroelectric order. The enhancement happens in regions with finite flexoelectric coupling and vanishing piezoelectricity, while the suppression happens in regions of constant strain. This is in good agreement with the trends observed in the RA-SHG measurements.

We finally estimate the ferroelectric transition temperature and its modification by the cavity from the temperature dependence of the steady-state FES mean-square displacement. 
The results in Fig.~6d show similar transition temperature with and without the cavity, with no statistically resolved shift of $T_c$.
This qualitatively agrees with the experimental observation that the ferroelectric transition temperature is unchanged by the cavity.
Although the simulations overestimate the experimental $T_c$, this is likely due to the fact that the macroscopic framework does not capture quantum fluctuations of the FES mode.

\noindent\textit{Conclusions}

Taken altogether, our work establishes that coupling to a dark THz-frequency cavity can induce, enhance, and suppress ferroelectric order in \STO. These effects stem from the ultrastrong  coupling between the cavity mode and the FES. From a careful analysis of the relationship between the observed cavity-induced structural response and the topographic inhomogeneity of the free-standing membranes, we conclude that this bi-directional control over the ferroelectric order stems from a difference in how the cavity influences regions of constant strain and regions with finite strain gradients. Together with our simulations, these observations highlight the role of local flexoelectric fields in the cavity-induced reorganization of the collective polarization profile, whose response is governed by fluctuations, nonlinearities, and spatial stiffness. More broadly, our findings show that spatially averaged observables can conceal substantial and opposing local cavity responses, underscoring the importance of spatially resolved measurements.

This work establishes several new avenues for cavity-based materials engineering. Not only have we demonstrated bi-directional cavity control over a long-range order for the first time, but we have also isolated a physical mechanism that determines whether enhancement or suppression is observed. As such, this work can be expanded to create photonic devices in which strain fields are engineered to drive specific desired phenomena. Rationally determining the response of a material to a cavity field creates the possibility to design a new generation of devices that leverage control over long-range collective phenomena of ordered phases rather than coupling of single quanta, as is typical of cavity QED devices. These possibilities range from driving novel photonic nonlinearities through cavity-engineering ferroelectricity, to designing novel THz-frequency quantum transducers that harness the cavity-induced structural phase transition, to quantum sensing devices that exploit the extreme sensitivities of these devices near the critical point. 

\newpage
\noindent\textbf{Methods}

\noindent\textit{Molecular Beam Epitaxy Synthesis of \STO}

All \STO\ samples were grown in a Veeco GEN10 molecular beam epitaxy (MBE) system.  Strontium and aluminum were supplied using conventional differentially pumped effusion cells, and titanium by a Veeco Ti-Ball source. The oxidant used was molecular oxygen at a background pressure of $1\times10^{-6}$ Torr. Films were grown on \STO\ (001) substrates using co-deposition for the Sr$_{4}$Al$_{2}$O$_{7}$ growth, and shuttered deposition for the \STO\ growth. All Sr$_{4}$Al$_{2}$O$_{7}$ layers were grown to be between 19 nm and 26 nm thick.

The samples were heated by an Epiray GmbH THERMALAS 1 kW CO$_{2}$ laser substrate heater, which has a wavelength of 10.6 $\mu$m and irradiates a circular area with a diameter of approximately 14 mm. The substrate temperature $T_{sub}$ was measured during growth by a pyrometer operating at 7.5 $\mu$m imaging the backside of the substrate. $T_{sub}$ was set to 700 $^{\circ}$C for both the \STO\ and Sr$_{4}$Al$_{2}$O$_{7}$ growths. After growth, the films were cooled to $T_{sub} < 300$ $^{\circ}$C in the same oxidant and pressure in which they were grown. Due to the excellent absorption of \STO\ at 10.6 $\mu$m, the backsides of the substrates were not coated. The stoichiometries, crystalline phases, and thicknesses of the \STO\ and Sr$_{4}$Al$_{2}$O$_{7}$ layes were confirmed using x-ray diffraction (XRD) analysis, which was was performed with a Panalytical Empyrean x-ray diffractometer using Cu K$\alpha$1 radiation. \\


\noindent\textit{\STO\ Membrane and Cavity Device Fabrication}

The first step in fabricating the \STO-PCC devices was forming the \STO\ membranes. PMMA (50K MW, 8\% in anisole by weight) was first spin coated (500 RPM/s for 5 s, 1000 RPM/s for 60 s) onto the surface of the \STO-Sr$_{4}$Al$_{2}$O$_{7}$ thin film heterostructure grown via MBE. The sample was then baked at 170 $^{\circ}$C for 2 min. The 10 mm $\times$ 10 mm substrates were then diced into 4 mm $\times$ 4 mm pieces. Each piece was then attached to a polymer stamp consisting of a PDMS layer coated with propylene carbonate (18\% in anisole by weight) via spin coating (750 RPM/s for 10 s, 2000 RPM/s for 60 s). After the attaching the sample, the stamp was baked for $\sim$5 minutes. After the stamp was prepared with the sample, the Sr$_{4}$Al$_{2}$O$_{7}$ was dissolved in deionized water and the freed substrate was removed. 

The stamp was then used to transfer the \STO\ membrane to the target substrate, such as the defect layer in the PCC devices. The transfer was performed by attaching the stamp to the substrate at 60 $^{\circ}$C, then the pair were heated to 225 $^{\circ}$C and held for 2 minutes. The stamp was then removed while the substrate was still at temperature, after which the substrate was cooled to below 60 $^{\circ}$C. The polymer residue was then removed via soaks in chloroform (45 min, room temperature), acetone (30 min, 60 $^{\circ}$C), and isopropyl alcohol (5 min).

The photonic crystal (PC) and photonic crystal cavity (PCC) devices were created by physically stacking alternate layers of diamond diamond and diamond rings that formed the air layers of the photonic structure. The rings and disks ranged from 10 $\mu$m thick to 50 $\mu$m thick, depending on the frequency of the cavity mode. The outer diameters of the rings and disks were 5 mm, and the inner diameters of the rings were 2.5 mm. The devices were stacked and held in custom sample holders machined out of copper. \\

\noindent\textit{Control measurements of bare versus cavity-controlled \STO\ response}

To ensure that the SHG response is quantitatively comparable between the cavity-coupled and bare samples, we fabricated several device geometries and verified that the results were consistent across them. Two types of devices were made, each with different normalization procedures. For the first type, the central defect layer was twice the thickness of the outer layers, as shown in Figure 1c. For these types of devices, the bare sample was accessed by removing the outer diamond layers. For SHG measurements, the SHG intensity was normalized between the bare and cavity-coupled configurations by correcting the cavity-coupled data for Fresnel reflection losses at the outer diamond layers, for both the incoming fundamental and the outgoing second-harmonic light. This correction was purely a multiplicative factor on the intensity; no other changes were made. This normalization procedure applied to the device shown in Figures 4 and 5. 

The second device geometry allows for control of the cavity mode while maintaining an identical light path between the sample and the measurement apparatus.  Instead of switching between the bare sample and the coupled sample by removing the DBR layers around the central defect layer, we instead split the defect layer into two separate layers---Layers A and B---that are the same thickness as the rest of the diamond layers in the PC. As shown in Extended Data Figure 6, Layer A serves as the substrate for the membrane. When Layer B is left out of the structure, there is no cavity resonance as the device is a perfect PC with no defect (Extended Data Figures 6a,b). When Layer B is inserted into the structure, the central layer becomes a defect, creating a sharp resonance (Extended Data Figures 6c,d). Importantly, the light path reflected from the \STO\ is identical, allowing for a controlled measurement comparing the coupled and uncoupled responses of the membrane without accounting for differences in the Fresnel reflections from the outer diamond layers. 

In both cases \textit{in situ} white-light imaging and pizeo-based micro-positioning is used to ensure that we are measuring the same spot on the sample in both the cavity and bare cases with below to 1 $\mu$m precision.

\noindent\textit{Time-Domain THz Spectroscopy Measurements}

The TDTS setup was seeded by 800 nm, 35 fs pulses produced by a Ti:sapphire amplifier operating at 1 kHz. 0.6 mJ of power was used to pump a spintronic THz emitter, and the residual 800 nm light was blocked by a high-resistivity float-zone silicon wafer. Detection of the THz electric field in the time-domain $E(t)$ was performed by electro-optic sampling (EOS) in a $<$110$>$-oriented ZnTe crystal with a thickness of 0.5 mm. The gate pulse used for EOS had $\sim$1 $\mu$J of pulse energy and was delayed in time with respect to the THz pulse using a mechanical delay stage. All temperature-dependent measurements were performed in a closed-cycle optical cryostat with an accessible range of 10 K to 300 K. 

The THz transmission spectra shown in the manuscript were performed by first measuring the THz electric after passing through the sample mounted on an aperture $E_{samp}(t)$ and then immediately measuring the THz electric field through an identical reference aperture $E_{ref}(t)$. Fast Fourier transforms of the $E(t)$ were performed to obtain the electric field as a function of frequency $\tilde{E}(\omega)$. The transmission $\tilde{t}(\omega)$ of the sample was then calculated as $\tilde{E}_{samp}(\omega)/\tilde{E}_{ref}(\omega)$. Plots of the transmission spectra are obtained by taking the absolute value $|\tilde{t}(\omega)|$. Movement between the two apertures was achieved via translation of the cryostat with a linear motorized stage.

To extract $\tilde{n}\left(\omega\right)$ from the TDTS data, the difference between the measured $\tilde{t}\left(\omega\right)$ and the expected theoretical response $\tilde{t}_{Th.}\left(\omega\right)$ was minimized using a least-squares algorithm at each $\omega$ with $\tilde{n}\left(\omega\right)$ as the variable of interest. $\tilde{t}_{Th.}\left(\omega\right)$ for a free-standing membrane on a substrate with index of refraction $n_{2}(\omega)$ is given by: 

\begin{equation}
\frac{2\tilde{n}(\omega)\left(n_{2}(\omega)+1\right)}{(\tilde{n}(\omega)+1)(\tilde{n}(\omega)+n_{2}(\omega))}\times\exp{(i\omega d(\tilde{n}(\omega)-1)/c)}\times FP(\tilde{n}(\omega))
\end{equation}

\noindent where $d$ is the sample thickness, $c$ is the speed of light in vacuum, and $FP(\tilde{n}(\omega))$ is the Fabry-P\'erot term that accounts for all internal reflections in the membrane: 

\begin{equation}
\frac{1}{1-\left(\frac{\tilde{n}(\omega)-1}{\tilde{n}(\omega)+1}\right)\left(\frac{\tilde{n}(\omega)-n_{2}(\omega)}{\tilde{n}(\omega)+n_{2}(\omega)}\right)}\times \exp(2i(\omega d/c)\tilde{n}(\omega)).
\end{equation}

\noindent After extraction of $\tilde{n}\left(\omega\right)$, all other optical constants such as the dielectric function $\tilde{\varepsilon}(\omega)$ can be calculated.

Temperature-dependent transmission measurements of the STO-PCC devices were normalized by the trace of a particular temperature within the series. This was done to isolate polaritonic effects and remove the background shape created by the photonic bandgap. The temperature chosen for normalization in each measurement is described in the main text. \\

\noindent\textit{Transfer Matrix Simulations}

Transfer matrix simulations were performed using the Near Field Optics package \cite{McLeod2017NearFieldOptics}. For the simulations in Figues 2c and 3c, a multilayer model was built to replicate the experimental device. For Figure 2c, this model included the following layers: diamond (23.0 $\mu$m), air (51.0 $\mu$m), diamond (43.0 $\mu$m), \STO\ (37 nm), air (51.0 $\mu$m)), diamond (23.0 $\mu$m). For each diamond layer, $\varepsilon_{diamond}(\omega)$ was set to 2.37. For the \STO\ layer, $\tilde{\epsilon}(\omega)$ was set to the experimental value extracted from the TDTS measurements shown in Extended Data Figure 3. Before feeding these experimental $\tilde{\epsilon}(\omega)$ into the simulation, they were first fitted with with a single Lorentizian: 

\begin{equation}
\varepsilon_{\infty} + A \frac{\omega_{0}^{2}}{\omega_{0}^{2} - \omega^{2} - ig\omega}
\end{equation}

\noindent where $A$ and $g$ are the strength and width of the Lorentzian, respectively, $\varepsilon_{\infty}$ is the term that accounts for higher-frequency contributions to the dielectric function that affect the low energy spectrum (Extended Data Figure 3). For Extended Data Figure 1, the simulation was identical to that done for Figure 2c, but the thicknesses of the \STO\ layers were adjusted to be 40 nm and 4 $\mu$m for Extended Data Figure 1a and Extended Data Figure 1b, respectively. The 140 K $\tilde{\varepsilon}(\omega)$ was used for the \STO\ layer. 

For the simulation in Figure 3c, the model included the following layers: diamond (12.0 $\mu$m), air (28.0 $\mu$m), diamond (12.0 $\mu$m), air (28.0 $\mu$m), diamond (24.0 $\mu$m), \STO\ (17 nm), air (28.0 $\mu$m), diamond (12.0 $\mu$m), air (28.0 $\mu$m), diamond (12.0 $\mu$m). For each diamond layer, $\varepsilon_{diamond}(\omega)$ was set to 2.37. For the \STO\ layer, $\tilde{\epsilon}(\omega)$ was set to the experimental value extracted from the TDTS measurements shown in Figure 3a. The experimental $\tilde{\epsilon}(\omega)$ data was smoothed before being fed into the simulation.\\

\noindent\textit{Rotational-Anisotropy Second Harmonic Generation Imaging}

The RA-SHG imaging setup was seeded by a femtosecond oscillator operating at 800 nm with a pulse width of 140 fs. The setup operated at normal incidence. Wide-field illumination of the sample was achieved by focusing 800 nm light on the back focal plane of a objective (Olympus, 0.45 NA, 20x magnification). Incoming light polarization was controlled with a polarizer-waveplate pair. Outgoing second-harmonic light was filtered with a dichroic mirror and a series of bandpass and short-pass filters. Outgoing polarization was selected with a polarizer that rotated in tandem with the waveplate on the incoming arm to produce the RA-SHG images and patterns. Images were detected using an EM-CCD (Princeton Instruments ProEM-HS 512). Sample temperature was controlled using a closed-cycle optical microscopy cryostat equipped with piezo nano-positioners. A radiation shield fixed at $\sim$4 K was positioned above the sample. 

There are three post-processing steps which are required to extract meaningful data from this type of data set: background subtraction, cosmic ray removal, and image overlap. The background in our images is created by the dark current of the EMCCD that occurs even when the camera shutter is fully closed. Since this background is uniform, we use a region of $50\times50$ pixels in a corner of the EMCCD that is not illuminated by the SHG image to define a background value. This average is then subtracted from the image. This averaging is performed separately for every collected image. 

Cosmic ray removal is performed to remove the hot-spots randomly created by these high-energy particles impacting the EMCCD. This procedure begins by setting a threshold value larger than the maximum of the SHG signal but below that of the cosmcic ray hot spots. All pixels with a value higher than the threshold are then checked, and those with a value higher than the threshold are replaced with the median value of all of its neighboring pixels. 

Finally, frame-to-frame image alignment was performed to compensate for the SHG image moving on the EMCCD due to the rotation of the polarization optics. This alignment was done using a phase cross correlation algorithm from the scikit-image Python library \cite{scikit-image}.\\ 

\noindent\textit{Second Harmonic Generation Polarimetry Fitting}

In its low-temperature paraelectric phase, \STO\ is a tetragonal perovskite with the point group $D_{4h}$. Since this point group preserves inversion symmetry, the leading-order contribution to SHG is the electric quadrupole (EQ) term: 
\begin{equation}
\label{EQ-SHG}
P^{(2\omega)}_i =  \chi^{(EQ)}_{ijkl} E_j(\omega) \nabla_k E_l(\omega)
\end{equation}
where $P^{(2\omega)}_i$ is the field-induced polarization in the sample at the SHG frequency $2\omega$ and $\chi^{(EQ)}_{ijkl}$ is the rank-4 nonlinear optical susceptibility tensor, and $E^{(\omega)}_j$ are the incoming probe electric fields at the fundamental frequency $\omega$. In our normal incidence geometry, the electric field of the light is represented as a plane wave traveling along the z direction in the lab frame, with a variable polarization determined by the angle $\phi$ in the x-y plane, 
\begin{equation}
    \label{z-electric-field}
    \textbf{E} = E_0e^{i(kz-\omega t)}(\sin(\phi)\hat{x} + \cos(\phi)\hat{y}).
\end{equation}
With this electric field, the only nonzero component of the gradient in Equation \ref{EQ-SHG} is $k=z$. Calculating the EQ-induced $\textbf{P}_{2\omega}$ based on the nonzero elements of $\boldsymbol{\chi}^{(EQ)}$ for the $D_{4h}$ space group yields
\begin{align}
   \textbf{P}_{2\omega} = \begin{bmatrix}
        0 \\
        0 \\
        \chi_{zxzx}\sin^2(\phi) + \chi_{zxzx}\cos^2(\phi)
    \end{bmatrix}
\end{align}
Only the $x$ and $y$ components of the polarization can be measured in our normal incidence geometry, so the contribution from the EQ SHG of the tetragonal paraelectric phase of \STO\ will always be 0.

The onset of ferroelectricity breaks inversion symmetry, and thus electric dipole (ED) SHG becomes the leading contribution:
\begin{equation}
    P^{(2\omega)}_i = \chi^{(ED)}_{ijk}E^{(\omega)}_jE^{(\omega)}_k
\end{equation}
where $\chi^{(ED)}_{ijk}$ is the rank-3 nonlinear optical susceptibility tensor. The ferroelectric phase of \STO\ is described by the $C_{4v}$ point group. The ED SHG produced will be dependent on the orientation of the ferroelectric domain. Assuming that the $c$-axis is oriented normal to the surface and that the electric field is oriented as in Equation \ref{z-electric-field}, then $c$-axis oriented ferroelectric domains will produce no ED SHG. The $a$-axis oriented tetragonal domains will produce the following $\textbf{P}_{2\omega}$:
\begin{align}
   \textbf{P}_{2\omega}(\phi) = E_{0}^{2}\begin{bmatrix}
        d_{31}\cos^2\phi + d_{33}\sin^2\phi \\
        d_{15}\cos\phi\sin\phi \\
        0
    \end{bmatrix}.
\end{align}
$b$-axis oriented domains will produce the following $\textbf{P}_{2\omega}$: 
\begin{align}
   \textbf{P}_{2\omega}(\phi) = E_{0}^{2}\begin{bmatrix}
        d_{31}\sin^2\phi + d_{33}\cos^2\phi \\
        d_{15}\cos\phi\sin\phi \\
        0
    \end{bmatrix}.
\end{align}
Since we measure $I_{2\omega}(\phi)\propto\left|\textbf{P}_{2\omega}(\phi)\right|^{2}$, we obtain the following functional form for both domains: 
\begin{equation}
\label{SHG_supp_final}
I_{2\omega}(\phi) = \epsilon_0E_0^2\left|A_1\cos^2(\phi) + A_2\sin^2(\phi) + B\cos(\phi)\sin(\phi)\right|^{2}.
\end{equation}
This form can yield either two- or four-fold rotationally-symmetric RA-SHG patterns depending on the relative values of $A_1$, $A_2$, and $B$:
\begin{itemize}
    \item{$A_1 = A_2 = 0$: Four-fold rotationally-symmetric pattern with four lobes offset by $45^{\circ}$ from the crystal axes.}
    \item{$A_1 = -A_2$: Four-fold rotationally-symmetric pattern with four lobes aligned with the crystal axes. Finite $B$ rotates the pattern away from the crystal axes.}
    \item{Only $A_{1}\neq0$ or $A_{2}\neq0$: Two-fold symmetric pattern with two lobes aligned along $0^{\circ}$ ($A_{1}\neq0$) or $90^{\circ}$ ($A_{2}\neq0$). Finite $B$ rotates the pattern away from the crystal axes.}
    \item{Both $A_{1}\neq0$ and $A_{2}\neq0$: If  $|A_{1}| > |A_{2}|$, two-fold symmetric pattern aligned along $0^{\circ}$. If  $|A_{1}| < |A_{2}|$, two-fold symmetric pattern aligned along $90^{\circ}$. Secondary lobes or absence of nodes are possible depending on relative values of $A_{1}$ and $A_{2}$. Finite $B$ rotates the pattern away from the crystal axes.}
\end{itemize}
We note that all possible sub-groups of the $D_{4h}$ point group will produce the same functional form as in Equation \ref{SHG_supp_final}. 

From this phenomenology, we can define a symmetry function:
\begin{equation}
    S = \left|A_{1}\right| -  \left|A_{2}\right|
\end{equation}
that allows us to characterize the degree to which a polar plot has four-fold symmetry and to classify the domain of the probed region. Patterns with $S=0$ will be four-fold rotationally symmetric, whereas patterns with $S>0$ will be in the $0^{\circ}$ domain and patterns with $S<0$ will be in the $90^{\circ}$ domain. We note that we never observe any four-fold symmetric patterns that are offset $45^{\circ}$ from the crystal axes, and so all $\bar{S}=0$ domains can be identified as the $A_{1} = -A_{2}$ type. We also note that regions with no polarization - i.e. a paraelectric domain - would also be classified as $S=0$.

In Figure 5 and Extended Data Figure 7, we use $S$ to create a spatial domain map of an RA-SHG image. We bin the image into $5\times5$-pixel $\left(\sim 0.7 \mu\text{m}\times0.7 \mu\text{m}\right)$ regions and fit the RA-SHG polar plot in each region with the formula in Equation \ref{SHG_supp_final}. 
To perform the fitting for of the binned regions, we first perform a normalization of the signal:
\begin{equation}
    I_{\rm norm}(\phi) = \frac{I(\phi) - I_{\rm min}}{I_{\rm max}}
\end{equation}
so that the maximum intensity of the signal will be approximately 1. Then for each of these regions the $S$ are calculated, allowing different domains to be mapped.

We note that RA-SHG traces were rotated by an offset before fitting to make the major lobes of the polar plots align with 0$^\circ$ or 90$^\circ$. This rotation causes the value of $B$ to be very small when measured in horizontal polarization, and was added because the offset between the initial polarization and the crystalline axes is not known with certainty due to the device geometry. We can thus not make a comment on the absolute value of the tensor elements of $\chi^{(ED)}_{ijk}$ because any information contained in the $B$ parameter is lost due to the manual rotation. Importantly, however, this rotation does not affect the spatial domain maps shown in Figure 5 and Extended Data Figure 7, as the fit still phenomenologically captures the symmetry of each SHG pattern via $A_1$ and $A_2$ regardless of the value of $B$, allowing us to create a spatial map based off of the $S$ parameter. \\

\noindent\textit{Microscopic multiscale simulations}

We model SrTiO$_3$ as a one-dimensional chain of $N$ unit cells along the polarization direction. Each cell hosts a ferroelectric soft (FES) mode $Q_f$ coupled to a lattice strain $u$, and the inter-cell coupling comprises dipole--dipole, short-range, and elastic-strain interactions. A single cavity mode couples collectively to the FES modes of all cells, and a flexoelectric term linking $Q_f$ to the strain gradient $\partial_x u$ breaks inversion symmetry. 
All on-site and inter-site potential parameters, as well as the flexoelectric coupling coefficient, are extracted from density functional theory (DFT) total-energy calculations of SrTiO$_3$ using the QUANTUM ESPRESSO package with the PBE functional \cite{Giannozzi2009,Perdew1996}. 
A detailed description of the DFT calculations and the evaluation of the flexoelectric coefficient are provided in the SI.

To describe the light--matter interaction, we employed the $N$-cell and single-photon effective Hamiltonian given by~\cite{Latini2021, Svendsen2025, I-Te2025}:
\begin{equation}
\begin{split}
\hat{H}_v &= \tfrac{1}{2}M_c\omega_c^2\,\hat{q}_c^2 + \frac{\hat{p}_c^2}{2M_c} + \sum_{i=1}^{N}\frac{\hat{p}_{\mathrm{u},i}^2}{2M_{\mathrm{u}}} + \sum_{i=1}^{N}\frac{1}{2M_f}\big[\hat{p}_{f,i} - g_c\hat{q}_c\big]^2 \\ 
&\quad + V_{\mathrm{DFT}}\!\left(\{\hat{u}\},\{\hat{Q}_{f}\}\right) + f_{\mathrm{flexo}}\sum_{i=1}^{N}\hat{Q}_{f,i}\, \partial_x\!\left(\frac{\hat{u}_{i}}{a_0}\right)
\end{split}
\label{eq:Hv}
\end{equation}
where $q_c$, $p_c$, $M_c$, $g_c$ are the photon coordinate, photon momentum, photon auxiliary mass, and light-matter coupling strength.
The potential $V_{\mathrm{DFT}}$ is the first-principles potential energy surface between on- and inter-site interactions. 
The flexoelectric coupling is employed to describe the local strain gradient-induced dipole.
The detailed description is in the SI.

Quantum fluctuations of both the FES mode and the cavity photon are captured using ring-polymer molecular dynamics (TRPMD) with the bare cavity frequency $\omega_c=1.5$ THz and coupled to a thermostat. 
In the high-frequency (adiabatic) limit, eliminating the photon dresses the collective ($\Gamma$) FES mode with an enhanced effective mass;
\begin{equation}
M_{\Gamma,\rm eff}=M_f\frac{\tilde{\omega}_c^2}{\omega_c^2}=M_f\;\left[1+\frac{N(\sqrt2A_0Z_f)^2}{M_cM_f\omega_c^2}\right]
\end{equation}
where $\tilde{\omega}_c^2=\omega_c^2(1+\eta^2)$ is the dressed cavity frequency. The cavity field thus acts to localize the FES mode distribution, as discussed in Ref.~\cite{Latini2021}.\\

\noindent\textit{Macroscopic simulations}

To describe the ferroelectric properties of STO over mesoscopic length scales, we derive an effective field theory from the microscopic model (see SI). Within the effective theory, the phonons are described by a continuum action $S[Q] = \int d\bx\, \mathcal{L}[Q]$, with the Lagrangian density given by
\begin{align}\label{eq:main_lagrangian}
 \mathcal{L} &= \frac{k_0}{2} (\partial_{t} Q)^2 - \frac{k_1}{2} (\partial_{\mathbf x} Q)^2 - \frac{k_2}{2} Q^2 - \frac{k_4}{4} Q^4 - \big(g_2 Q^2 + g_4 Q^4 \big) u + \frac{f_1}{2} \big( Q \partial_{x} u - u \partial_{x} Q\big).
\end{align}
Here, the term proportional to $k_0$ is the kinetic energy, $k_1$ describes the cost of spatial fluctuations, $k_2$ is a spring constant, and $k_4$ is the leading non-linearity. Together, these terms describe the internal non-linear dynamics of the phonon mode, while the terms proportional to $g_i$ and $f_i$ describe its piezo- and flexoelectric coupling to strain and strain gradients, respectively.

To simulate the phonon dynamics, we solve the equations of motion derived from this Lagrangian, promoted to stochastic equations by including Gaussian white noise. This allows to simulate the thermal state of the system. The resulting phonon equation of motion is a non-linear stochastic wave equation of the form
\begin{align}\label{eq:main_eom}
 \big[k_0 \partial_{t}^2 - k_1 \partial_{\mathbf x}^2 - k_2 + \gamma \partial_t \big] Q = k_5 Q^3 - \big( 2g_2 Q + 4g_4 Q^3 \big) u + f_1 \partial_{x} u  + \xi(\bx,t), \nonumber
\end{align}
where $\gamma$ is a phenomenological damping and $\xi$ accounts for the stochastic noise term. Simulations were performed in a rectangular domain of dimensions $D = [0, L_x] \times [0, L_y]$, with each axis discretized into a grid of $N_x$ and $N_y$ points, respectively. We numerically propagate the equations of motion inside the time interval $T = [t_0, t_1]$, discretized into $N_t$ time steps, employing periodic boundary conditions in both spatial directions.

In all calculations, we start from a random initial state where the coordinates $Q(\bx_i)$ and velocities $V(\bx_i)$ are drawn from the uniform distribution $U(-1,1)$. The system is then propagated until it reaches a steady state. In practice, we found it sufficient to use $N_t = 20000$ and a time step of $dt = 0.05$. For the results of Fig.~6, we used $N_x = 600$ and $N_y = 360$, amounting to a total number of grid points $N_x N_y = 216000$, where each grid point corresponds to a region $10 \times 10$ nm$^2$ in size. All parameters used in the simulations are reported in Tab.~S1 of the SI. To describe the effect of the cavity on the ferroelectric soft mode, we increase the effective phonon mass as discussed above. To obtain expectation values like $\langle Q({\mathbf r})\rangle$, we sample over $N_{\mathrm tr} = 500$ independent trajectories, and calculate the statistical average.

\newpage

\noindent\textbf{Acknowledgments}
\noindent The authors thank Valla Fatemi for useful discussions. We thank Xiaoxi Huang for assistance with spintronic THz emitter fabrication. We thank Shivasheesh Varsheny and Bharat Jalan for providing preliminary \STO\ membranes for testing. We thank Michele Buzzi for assisting with the design of the PCC. This work is supported by DARPA Contract No. HR0011-25-3-0106. This work made use of the synthesis facilities of the Platform for the Accelerated Realization, Analysis, and Discovery of Interface Materials (PARADIM), which are supported by the National Science Foundation (NSF) under Cooperative Agreement No. DMR-2039380. This work was performed in part at the Cornell NanoScale Facility, an NNCI member supported by NSF Grant NNCI-2025233. O.M. and A.T.P. acknowledge support from the Kavli Institute at Cornell (KIC) Postdoctoral Fellowship.
H.K. and D.S. were supported by the National Research Foundation of Korea (NRF) grant funded by the Korea government (MSIT) (No. RS-2024-00333664 and RS-2026-25508298)\\

\noindent\textbf{Author Contributions Statement}

\noindent O.M. and A.S.D. conceived the experiment and designed the device architecture. O.M., S.G, D.P.S., and D.G.S. grew the \STO\ membranes. O.M. and S.G. fabricated the cavity devices. O.M. and S.G. performed the time-domain THz spectroscopy measurements. O.M., S.G., and A.T.P. performed the rotational-anisotropy second harmonic generation imaging experiments. O.M. performed the transfer matrix simulations. O.M. and S.G. performed all data analysis. The DFT-based multiscale simulations were performed by H.K., S.L., D.S., and A.R. The Ginzburg-Landau calculations were performed by E.V.B and A.R. All authors discussed the experimental findings and interpreted the results. O.M., S.G., and A.S.D. wrote the manuscript with input from all authors.\\

\noindent\textbf{Competing Interests Statement} 

\noindent The authors declare no competing interests.\\

\noindent\textbf{Data Availability}

\noindent All data from the the plots within this paper and other findings of this study are available from the corresponding authors upon reasonable request.\\

\noindent\textbf{Code Availability}

\noindent All code from the the plots within this paper and other findings of this study are available from the corresponding authors upon reasonable request.

\bibliography{References_main}

@misc{McLeod2017NearFieldOptics,
  author       = {McLeod, Alexander S.},
  title        = {{NearFieldOptics}: Tools for computing optical properties of materials and measurable quantities for near-field microscopy},
  year         = {2017},
  howpublished = {\url{https://github.com/asmcleod/NearFieldOptics}},
  note         = {GitHub repository, MIT License.}
}

@misc{Montanaro_2026,
  author        = {Montanaro, Angela and Plastovets, Vadim and Khatiwada, Nitesh and Fiore, Jacopo and Jarc, Giacomo and Alabbadi, Abdullah and Mastropasqua, Antonio and Rigoni, Enrico Maria and Mathengattil, Shahla Y. and Dal Zilio, Simone and Fassioli Olsen, Francesca and Novelli, Fabio and Winnerl, Stephan and Sentef, Michael A. and Kennes, Dante M. and Millis, Andrew J. and Piazza, Francesco and Fausti, Daniele},
  title         = {Cavity-enhanced superconducting response in an underdoped cuprate},
  year          = {2026},
  eprint        = {2606.18084},
  archivePrefix = {arXiv},
  primaryClass  = {cond-mat.supr-con},
}

@misc{Zhang_NbSe2_2026,
  author        = {Zhang, Hanxiang and Feng, Zexin and Lu, I-Te and Li, Zhiwei and Guo, Songhao and Shang, Qiuyu and Tan, Thomas and Lyu, Xiaodan and Shen, Xiangming and Luan, Dening and Panmai, Mingcheng and Watanabe, Kenji and Taniguchi, Takashi and Singh, Ranjan and Rubio, Angel and Gao, Weibo},
  title         = {Cavity-enhanced superconductivity in the two-dimensional limit of {NbSe$_2$}},
  year          = {2026},
  eprint        = {2606.19171},
  archivePrefix = {arXiv},
  primaryClass  = {cond-mat.supr-con},
}

@article{frisk_kockum_ultrastrong_2019,
	title = {Ultrastrong coupling between light and matter},
	volume = {1},
	copyright = {2019 Springer Nature Limited},
	issn = {2522-5820},
	doi = {10.1038/s42254-018-0006-2},
	language = {en},
	urldate = {2022-12-15},
	journal = {Nat. Rev. Phys.},
	publisher = {Nature Publishing Group},
	author = {Frisk Kockum, Anton and Miranowicz, Adam and De Liberato, Simone and Savasta, Salvatore and Nori, Franco},
	month = jan,
	year = {2019},
	pages = {19--40},
}

@article{brennecke_cavity_2007,
	title = {Cavity {QED} with a {Bose}–{Einstein} condensate},
	volume = {450},
	copyright = {2007 Springer Nature Limited},
	issn = {1476-4687},
	doi = {10.1038/nature06120},
	language = {en},
	number = {7167},
	urldate = {2026-07-30},
	journal = {Nature},
	publisher = {Nature Publishing Group},
	author = {Brennecke, Ferdinand and Donner, Tobias and Ritter, Stephan and Bourdel, Thomas and Köhl, Michael and Esslinger, Tilman},
	month = nov,
	year = {2007},
	pages = {268--271},
}

@article{sauer_cavity_2004,
	title = {Cavity {QED} with optically transported atoms},
	volume = {69},
	doi = {10.1103/PhysRevA.69.051804},
	number = {5},
	urldate = {2026-07-30},
	journal = {Phys. Rev. A},
	publisher = {American Physical Society},
	author = {Sauer, J. A. and Fortier, K. M. and Chang, M. S. and Hamley, C. D. and Chapman, M. S.},
	month = may,
	year = {2004},
	pages = {051804},
}

@article{cao_qubit_2011,
	title = {A qubit strongly coupled to a resonant cavity: asymmetry of the spontaneous emission spectrum beyond the rotating wave approximation},
	volume = {13},
	issn = {1367-2630},
	shorttitle = {A qubit strongly coupled to a resonant cavity},
	doi = {10.1088/1367-2630/13/7/073002},
	language = {en},
	number = {7},
	urldate = {2026-07-30},
	journal = {New J. Phys.},
	author = {Cao, X and You, J Q and Zheng, H and Nori, F},
	month = jul,
	year = {2011},
	pages = {073002},
}

@article{ciuti_quantum_2005,
	title = {Quantum vacuum properties of the intersubband cavity polariton field},
	volume = {72},
	doi = {10.1103/PhysRevB.72.115303},
	number = {11},
	urldate = {2026-07-30},
	journal = {Phys. Rev. B},
	publisher = {American Physical Society},
	author = {Ciuti, Cristiano and Bastard, Gérald and Carusotto, Iacopo},
	month = sep,
	year = {2005},
	pages = {115303},
}

@article{wang_phase_1973,
	title = {Phase {Transition} in the {Dicke} {Model} of {Superradiance}},
	volume = {7},
	doi = {10.1103/PhysRevA.7.831},
	number = {3},
	urldate = {2026-07-30},
	journal = {Phys. Rev. A},
	publisher = {American Physical Society},
	author = {Wang, Y. K. and Hioe, F. T.},
	month = mar,
	year = {1973},
	pages = {831--836},
}

@article{de_liberato_light-matter_2014,
	title = {Light-{Matter} {Decoupling} in the {Deep} {Strong} {Coupling} {Regime}: {The} {Breakdown} of the {Purcell} {Effect}},
	volume = {112},
	shorttitle = {Light-{Matter} {Decoupling} in the {Deep} {Strong} {Coupling} {Regime}},
	doi = {10.1103/PhysRevLett.112.016401},
	number = {1},
	urldate = {2026-07-30},
	journal = {Phys. Rev. Lett.},
	publisher = {American Physical Society},
	author = {De Liberato, Simone},
	month = jan,
	year = {2014},
	pages = {016401},
}

@article{schoenlein_femtosecond_1987,
	title = {Femtosecond studies of nonequilibrium electronic processes in metals},
	volume = {58},
	doi = {10.1103/PhysRevLett.58.1680},
	number = {16},
	urldate = {2026-07-30},
	journal = {Phys. Rev. Lett.},
	publisher = {American Physical Society},
	author = {Schoenlein, R. W. and Lin, W. Z. and Fujimoto, J. G. and Eesley, G. L.},
	month = apr,
	year = {1987},
	pages = {1680--1683},
}

@article{wall_ultrafast_2012,
	title = {Ultrafast changes in lattice symmetry probed by coherent phonons},
	volume = {3},
	copyright = {2012 Springer Nature Limited},
	issn = {2041-1723},
	doi = {10.1038/ncomms1719},
	language = {en},
	number = {1},
	urldate = {2026-07-30},
	journal = {Nat. Commun.},
	publisher = {Nature Publishing Group},
	author = {Wall, S. and Wegkamp, D. and Foglia, L. and Appavoo, K. and Nag, J. and Haglund, R. F. and Stähler, J. and Wolf, M.},
	month = mar,
	year = {2012},
	pages = {721},
}

@article{wang_observation_2013,
	title = {Observation of {Floquet}-{Bloch} {States} on the {Surface} of a {Topological} {Insulator}},
	volume = {342},
	doi = {10.1126/science.1239834},
	number = {6157},
	urldate = {2026-07-30},
	journal = {Science},
	publisher = {American Association for the Advancement of Science},
	author = {Wang, Y. H. and Steinberg, H. and Jarillo-Herrero, P. and Gedik, N.},
	month = oct,
	year = {2013},
	pages = {453--457},
}

@article{mikhaylovskiy_ultrafast_2015,
	title = {Ultrafast optical modification of exchange interactions in iron oxides},
	volume = {6},
	copyright = {2015 The Author(s)},
	issn = {2041-1723},
	doi = {10.1038/ncomms9190},
	language = {en},
	number = {1},
	urldate = {2026-07-30},
	journal = {Nat. Commun.},
	publisher = {Nature Publishing Group},
	author = {Mikhaylovskiy, R. V. and Hendry, E. and Secchi, A. and Mentink, J. H. and Eckstein, M. and Wu, A. and Pisarev, R. V. and Kruglyak, V. V. and Katsnelson, M. I. and Rasing, Th and Kimel, A. V.},
	month = sep,
	year = {2015},
	pages = {8190},
}

@article{juraschek_ultrafast_2017,
	title = {Ultrafast {Structure} {Switching} through {Nonlinear} {Phononics}},
	volume = {118},
	doi = {10.1103/PhysRevLett.118.054101},
	number = {5},
	urldate = {2026-07-29},
	journal = {Phys. Rev. Lett.},
	publisher = {American Physical Society},
	author = {Juraschek, D. M. and Fechner, M. and Spaldin, N. A.},
	month = jan,
	year = {2017},
	pages = {054101},
}

@article{fausti_light-induced_2011,
	title = {Light-{Induced} {Superconductivity} in a {Stripe}-{Ordered} {Cuprate}},
	volume = {331},
	issn = {0036-8075, 1095-9203},
	doi = {10.1126/science.1197294},
	language = {en},
	number = {6014},
	urldate = {2026-07-29},
	journal = {Science},
	author = {Fausti, D. and Tobey, R. I. and Dean, N. and Kaiser, S. and Dienst, A. and Hoffmann, M. C. and Pyon, S. and Takayama, T. and Takagi, H. and Cavalleri, A.},
	month = jan,
	year = {2011},
	pages = {189--191},
}

@article{li_keldysh_2022,
	title = {Keldysh {Space} {Control} of {Charge} {Dynamics} in a {Strongly} {Driven} {Mott} {Insulator}},
	language = {en},
	journal = {Phys. Rev. Lett.},
	author = {Li, Xinwei and Ning, Honglie and Mehio, Omar and Zhao, Hengdi and Lee, Min-Cheol and Kim, Kyungwan and Nakamura, Fumihiko and Maeno, Yoshiteru and Cao, Gang and Hsieh, David},
	year = {2022},
}

@article{denev_probing_2011,
	title = {Probing {Ferroelectrics} {Using} {Optical} {Second} {Harmonic} {Generation}},
	volume = {94},
	copyright = {© 2011 The American Ceramic Society},
	issn = {1551-2916},
	doi = {10.1111/j.1551-2916.2011.04740.x},
	language = {en},
	number = {9},
	urldate = {2026-07-29},
	journal = {Journal of the American Ceramic Society},
	author = {Denev, Sava A. and Lummen, Tom T. A. and Barnes, Eftihia and Kumar, Amit and Gopalan, Venkatraman},
	year = {2011},
	pages = {2699--2727},
}

@article{ashida_quantum_2020,
	title = {Quantum {Electrodynamic} {Control} of {Matter}: {Cavity}-{Enhanced} {Ferroelectric} {Phase} {Transition}},
	volume = {10},
	issn = {2160-3308},
	shorttitle = {Quantum {Electrodynamic} {Control} of {Matter}},
	doi = {10.1103/PhysRevX.10.041027},
	language = {en},
	number = {4},
	urldate = {2022-12-24},
	journal = {Phys. Rev. X},
	author = {Ashida, Yuto and İmamoğlu, Ataç and Faist, Jérôme and Jaksch, Dieter and Cavalleri, Andrea and Demler, Eugene},
	month = nov,
	year = {2020},
	pages = {041027},
}

@article{anappara_signatures_2009,
	title = {Signatures of the ultrastrong light-matter coupling regime},
	volume = {79},
	doi = {10.1103/PhysRevB.79.201303},
	number = {20},
	urldate = {2026-07-29},
	journal = {Phys. Rev. B},
	publisher = {American Physical Society},
	author = {Anappara, Aji A. and De Liberato, Simone and Tredicucci, Alessandro and Ciuti, Cristiano and Biasiol, Giorgio and Sorba, Lucia and Beltram, Fabio},
	month = may,
	year = {2009},
	pages = {201303},
}

@article{niemczyk_circuit_2010,
	title = {Circuit quantum electrodynamics in the ultrastrong-coupling regime},
	volume = {6},
	copyright = {2010 Springer Nature Limited},
	issn = {1745-2481},
	doi = {10.1038/nphys1730},
	language = {en},
	number = {10},
	urldate = {2026-07-29},
	journal = {Nat. Phys.},
	publisher = {Nature Publishing Group},
	author = {Niemczyk, T. and Deppe, F. and Huebl, H. and Menzel, E. P. and Hocke, F. and Schwarz, M. J. and Garcia-Ripoll, J. J. and Zueco, D. and Hümmer, T. and Solano, E. and Marx, A. and Gross, R.},
	month = oct,
	year = {2010},
	pages = {772--776},
}

@article{wang_cavity_2019,
	title = {Cavity quantum electrodynamical {Chern} insulator: {Towards} light-induced quantized anomalous {Hall} effect in graphene},
	volume = {99},
	shorttitle = {Cavity quantum electrodynamical {Chern} insulator},
	doi = {10.1103/PhysRevB.99.235156},
	number = {23},
	urldate = {2026-07-29},
	journal = {Phys. Rev. B},
	publisher = {American Physical Society},
	author = {Wang, Xiao and Ronca, Enrico and Sentef, Michael A.},
	month = jun,
	year = {2019},
	pages = {235156},
}

@article{nataf_rashba_2019,
	title = {Rashba {Cavity} {QED}: {A} {Route} {Towards} the {Superradiant} {Quantum} {Phase} {Transition}},
	volume = {123},
	shorttitle = {Rashba {Cavity} {QED}},
	doi = {10.1103/PhysRevLett.123.207402},
	number = {20},
	urldate = {2026-07-29},
	journal = {Phys. Rev. Lett.},
	publisher = {American Physical Society},
	author = {Nataf, Pierre and Champel, Thierry and Blatter, Gianni and Basko, Denis M.},
	month = nov,
	year = {2019},
	pages = {207402},
}

@article{lu_cavity-enhanced_2024,
	title = {Cavity-enhanced superconductivity in {MgB$_2$} from first-principles quantum electrodynamics ({QEDFT})},
	volume = {121},
	doi = {10.1073/pnas.2415061121},
	number = {50},
	urldate = {2025-03-18},
	journal = {Proceedings of the National Academy of Sciences},
	publisher = {Proceedings of the National Academy of Sciences},
	author = {Lu, I-Te and Shin, Dongbin and Svendsen, Mark Kamper and Hübener, Hannes and De Giovannini, Umberto and Latini, Simone and Ruggenthaler, Michael and Rubio, Angel},
	month = dec,
	year = {2024},
	pages = {e2415061121},
}

@article{nova_metastable_2019,
	title = {Metastable ferroelectricity in optically strained {SrTiO3}},
	volume = {364},
	doi = {10.1126/science.aaw4911},
	number = {6445},
	urldate = {2023-02-16},
	journal = {Science},
	publisher = {American Association for the Advancement of Science},
	author = {Nova, T. F. and Disa, A. S. and Fechner, M. and Cavalleri, A.},
	month = jun,
	year = {2019},
	pages = {1075--1079},
}

@article{floch_electromagnetic_2011,
	title = {Electromagnetic properties of polycrystalline diamond from 35 {K} to room temperature and microwave to terahertz frequencies},
	volume = {109},
	issn = {0021-8979},
	doi = {10.1063/1.3580903},
	number = {9},
	urldate = {2023-05-23},
	journal = {Journal of Applied Physics},
	author = {Floch, Jean-Michel Le and Bara, Romain and Hartnett, John G. and Tobar, Michael E. and Mouneyrac, David and Passerieux, Damien and Cros, Dominique and Krupka, Jerzy and Goy, Philippe and Caroopen, Sylvain},
	month = may,
	year = {2011},
	pages = {094103},
}

@article{lu_synthesis_2016,
	title = {Synthesis of freestanding single-crystal perovskite films and heterostructures by etching of sacrificial water-soluble layers},
	volume = {15},
	copyright = {2016 Springer Nature Limited},
	issn = {1476-4660},
	doi = {10.1038/nmat4749},
	language = {en},
	number = {12},
	urldate = {2023-06-21},
	journal = {Nature Mater.},
	publisher = {Nature Publishing Group},
	author = {Lu, Di and Baek, David J. and Hong, Seung Sae and Kourkoutis, Lena F. and Hikita, Yasuyuki and Hwang, Harold Y.},
	month = dec,
	year = {2016},
	note = {Number: 12},
	pages = {1255--1260},
}

@article{chiu_presence_2022,
	title = {Presence of {Delocalized} {Ti} 3d {Electrons} in {Ultrathin} {Single}-{Crystal} {SrTiO$_3$}},
	volume = {22},
	issn = {1530-6984},
	doi = {10.1021/acs.nanolett.1c04434},
	number = {4},
	urldate = {2024-08-29},
	journal = {Nano Lett.},
	publisher = {American Chemical Society},
	author = {Chiu, Chun-Chien and Ho, Sheng-Zhu and Lee, Jenn-Min and Shao, Yu-Cheng and Shen, Yang and Liu, Yu-Chen and Chang, Yao-Wen and Zheng, Yun-Zhe and Huang, Rong and Chang, Chun-Fu and Kuo, Chang-Yang and Duan, Chun-Gang and Huang, Shih-Wen and Yang, Jan-Chi and Chuang, Yi-De},
	month = feb,
	year = {2022},
	pages = {1580--1586},
}

@article{vinas_bostrom_controlling_2023,
	title = {Controlling the magnetic state of the proximate quantum spin liquid $\alpha$-{RuCl$_3$} with an optical cavity},
	volume = {9},
	copyright = {2023 The Author(s)},
	issn = {2057-3960},
	doi = {10.1038/s41524-023-01158-6},
	language = {en},
	number = {1},
	urldate = {2024-09-04},
	journal = {npj Comput. Mater.},
	publisher = {Nature Publishing Group},
	author = {Viñas Boström, Emil and Sriram, Adithya and Claassen, Martin and Rubio, Angel},
	month = oct,
	year = {2023},
	pages = {1--10},
}

@article{li_vacuum_2018,
	title = {Vacuum {Bloch}–{Siegert} shift in {Landau} polaritons with ultra-high cooperativity},
	volume = {12},
	copyright = {2018 The Author(s)},
	issn = {1749-4893},
	doi = {10.1038/s41566-018-0153-0},
	language = {en},
	number = {6},
	urldate = {2025-04-01},
	journal = {Nat. Photon.},
	publisher = {Nature Publishing Group},
	author = {Li, Xinwei and Bamba, Motoaki and Zhang, Qi and Fallahi, Saeed and Gardner, Geoff C. and Gao, Weilu and Lou, Minhan and Yoshioka, Katsumasa and Manfra, Michael J. and Kono, Junichiro},
	month = jun,
	year = {2018},
	pages = {324--329},
}

@article{li_classical--quantum_2025,
	title = {The classical-to-quantum crossover in the strain-induced ferroelectric transition in {SrTiO$_3$} membranes},
	volume = {16},
	copyright = {2025 The Author(s)},
	issn = {2041-1723},
	doi = {10.1038/s41467-025-59517-4},
	language = {en},
	number = {1},
	urldate = {2025-05-30},
	journal = {Nat. Commun.},
	publisher = {Nature Publishing Group},
	author = {Li, Jiarui and Lee, Yonghun and Choi, Yongseong and Kim, Jong-Woo and Thompson, Paul and Crust, Kevin J. and Xu, Ruijuan and Hwang, Harold Y. and Ryan, Philip J. and Lee, Wei-Sheng},
	month = may,
	year = {2025},
	pages = {4445},
}

@article{zhang_collective_2016,
	title = {Collective non-perturbative coupling of {2D} electrons with high-quality-factor terahertz cavity photons},
	volume = {12},
	language = {en},
	journal = {Nat. Phys.},
	author = {Zhang, Qi and Lou, Minhan and Li, Xinwei and Reno, John L and Pan, Wei and Watson, John D and Manfra, Michael J and Kono, Junichiro},
	year = {2016},
}

@article{forst_nonlinear_2011,
	title = {Nonlinear phononics as an ultrafast route to lattice control},
	volume = {7},
	copyright = {2011 Springer Nature Limited},
	issn = {1745-2481},
	doi = {10.1038/nphys2055},
	language = {en},
	number = {11},
	urldate = {2026-07-28},
	journal = {Nat. Phys.},
	publisher = {Nature Publishing Group},
	author = {Först, M. and Manzoni, C. and Kaiser, S. and Tomioka, Y. and Tokura, Y. and Merlin, R. and Cavalleri, A.},
	month = nov,
	year = {2011},
	pages = {854--856},
}

@article{graziotto_cavity_2026,
	title = {Cavity quantum electrodynamics control of quantum {Hall} stripes},
	copyright = {2026 The Author(s)},
	issn = {1745-2481},
	doi = {10.1038/s41567-026-03287-3},
	language = {en},
	urldate = {2026-07-28},
	journal = {Nat. Phys.},
	publisher = {Nature Publishing Group},
	author = {Graziotto, Lorenzo and Enkner, Josefine and Chattopadhyay, Sambuddha and Curtis, Jonathan B. and Koskas, Ethan and Reichl, Christian and Wegscheider, Werner and Scalari, Giacomo and Demler, Eugene and Faist, Jérôme},
	month = may,
	year = {2026},
	pages = {1--6},
}

@article{keren_cavity-altered_2026,
	title = {Cavity-altered superconductivity},
	volume = {650},
	copyright = {2026 The Author(s)},
	issn = {1476-4687},
	doi = {10.1038/s41586-025-10062-6},
	language = {en},
	number = {8103},
	urldate = {2026-07-28},
	journal = {Nature},
	publisher = {Nature Publishing Group},
	author = {Keren, Itai and Webb, Tatiana A. and Zhang, Shuai and Xu, Jikai and Sun, Dihao and Kim, Brian S. Y. and Shin, Dongbin and Zhang, Songtian S. and Zhang, Junhe and Pereira, Giancarlo and Yao, Juntao and Okugawa, Takuya and Michael, Marios H. and Viñas Boström, Emil and Edgar, James H. and Wolf, Stuart and Julian, Matthew and Prasankumar, Rohit P. and Miyagawa, Kazuya and Kanoda, Kazushi and Gu, Genda and Cothrine, Matthew and Mandrus, David and Buzzi, Michele and Cavalleri, Andrea and Dean, Cory R. and Kennes, Dante M. and Millis, Andrew J. and Li, Qiang and Sentef, Michael A. and Rubio, Angel and Pasupathy, Abhay N. and Basov, D. N.},
	month = feb,
	year = {2026},
	pages = {864--868},
}

@article{muller_srtimathrmo_3_1979,
	title = {{SrTiO$_3$}: {An} intrinsic quantum paraelectric below 4 {K}},
	volume = {19},
	shorttitle = {{SrTi}\$\{{\textbackslash}mathrm\{{O}\}\}\_\{3\}\$},
	doi = {10.1103/PhysRevB.19.3593},
	number = {7},
	urldate = {2026-07-28},
	journal = {Phys. Rev. B},
	publisher = {American Physical Society},
	author = {Müller, K. A. and Burkard, H.},
	month = apr,
	year = {1979},
	pages = {3593--3602},
}

@article{yang_epitaxial_2022,
	title = {Epitaxial {SrTiO$_3$} films with dielectric constants exceeding 25,000},
	volume = {119},
	doi = {10.1073/pnas.2202189119},
	number = {23},
	urldate = {2026-07-28},
	journal = {Proceedings of the National Academy of Sciences},
	publisher = {Proceedings of the National Academy of Sciences},
	author = {Yang, Zhifei and Lee, Dooyong and Yue, Jin and Gabel, Judith and Lee, Tien-Lin and James, Richard D. and Chambers, Scott A. and Jalan, Bharat},
	month = jun,
	year = {2022},
	pages = {e2202189119},
}

@article{sirenko_soft-mode_2000,
	title = {Soft-mode hardening in {SrTiO$_3$} thin films},
	volume = {404},
	copyright = {2000 Macmillan Magazines Ltd.},
	issn = {1476-4687},
	doi = {10.1038/35006023},
	language = {en},
	number = {6776},
	urldate = {2026-07-28},
	journal = {Nature},
	publisher = {Nature Publishing Group},
	author = {Sirenko, A. A. and Bernhard, C. and Golnik, A. and Clark, Anna M. and Hao, Jianhua and Si, Weidong and Xi, X. X.},
	month = mar,
	year = {2000},
	pages = {373--376},
}

@article{zhang_super-tetragonal_2024,
	title = {Super-tetragonal {Sr$_4$Al$_2$O$_7$} as a sacrificial layer for high-integrity freestanding oxide membranes},
	volume = {383},
	doi = {10.1126/science.adi6620},
	number = {6681},
	urldate = {2025-06-13},
	journal = {Science},
	publisher = {American Association for the Advancement of Science},
	author = {Zhang, Jinfeng and Lin, Ting and Wang, Ao and Wang, Xiaochao and He, Qingyu and Ye, Huan and Lu, Jingdi and Wang, Qing and Liang, Zhengguo and Jin, Feng and Chen, Shengru and Fan, Minghui and Guo, Er-Jia and Zhang, Qinghua and Gu, Lin and Luo, Zhenlin and Si, Liang and Wu, Wenbin and Wang, Lingfei},
	month = jan,
	year = {2024},
	pages = {388--394},
}

@article{stoupin_thermal_2010,
	title = {Thermal {Expansion} of {Diamond} at {Low} {Temperatures}},
	volume = {104},
	copyright = {http://link.aps.org/licenses/aps-default-license},
	issn = {0031-9007, 1079-7114},
	doi = {10.1103/PhysRevLett.104.085901},
	language = {en},
	number = {8},
	urldate = {2026-07-28},
	journal = {Phys. Rev. Lett.},
	author = {Stoupin, Stanislav and Shvyd’ko, Yuri V.},
	month = feb,
	year = {2010},
	pages = {085901},
}

@article{xiao_terahertz_2023,
	title = {Terahertz optoelectronic properties of synthetic single crystal diamond},
	volume = {139},
	issn = {0925-9635},
	doi = {10.1016/j.diamond.2023.110266},
	urldate = {2026-07-28},
	journal = {Diamond and Related Materials},
	author = {Xiao, Huan and Zhang, Zongyan and Xu, Wen and Wang, Qiujing and Xiao, Yiming and Ding, Lan and Huang, Jiangtao and Li, Haowen and He, Bin and Peeters, Francois M.},
	month = nov,
	year = {2023},
	pages = {110266},
}

@article{jarc_cavity-mediated_2023,
	title = {Cavity-mediated thermal control of metal-to-insulator transition in {1T}-{TaS$_2$}},
	volume = {622},
	copyright = {2023 The Author(s), under exclusive licence to Springer Nature Limited},
	issn = {1476-4687},
	doi = {10.1038/s41586-023-06596-2},
	language = {en},
	number = {7983},
	urldate = {2026-07-28},
	journal = {Nature},
	publisher = {Nature Publishing Group},
	author = {Jarc, Giacomo and Mathengattil, Shahla Yasmin and Montanaro, Angela and Giusti, Francesca and Rigoni, Enrico Maria and Sergo, Rudi and Fassioli, Francesca and Winnerl, Stephan and Dal Zilio, Simone and Mihailovic, Dragan and Prelovšek, Peter and Eckstein, Martin and Fausti, Daniele},
	month = oct,
	year = {2023},
	pages = {487--492},
}

@article{appugliese_breakdown_2022,
	title = {Breakdown of topological protection by cavity vacuum fields in the integer quantum {Hall} effect},
	volume = {375},
	doi = {10.1126/science.abl5818},
	language = {en},
	urldate = {2026-07-28},
	journal = {Science},
	author = {Appugliese, Felice and Enkner, Josefine and Paravicini-Bagliani, Gian Lorenzo and Beck, Mattias and Reichl, Christian and Wegscheider, Werner and Scalari, Giacomo and Ciuti, Cristiano and Faist, Jerome},
	month = mar,
	year = {2022},
	doi = {10.1126/science.abl5818},
	pages = {1030--1034},
}

@article{wang_evidence_2026,
	title = {Evidence for vacuum-enhanced superconductivity in {NbSe$_2$}},
	volume = {657},
	copyright = {2026 The Author(s), under exclusive licence to Springer Nature Limited},
	issn = {1476-4687},
	doi = {10.1038/s41586-026-11037-x},
	language = {en},
	number = {8133},
	urldate = {2026-10-05},
	journal = {Nature},
	publisher = {Nature Publishing Group},
	author = {Wang, Zheyan and Cardoso, Gabriel and Yang, Liu and Gong, Xun and Zhang, Chi and Zhu, Yufei and Zhang, Dongbo and Pan, Nan and Cai, Hongbing and Chen, Yong P. and Jiang, Qing-Dong and Cheng, Guanghui and Wilczek, Frank and Zeng, Changgan},
	month = sep,
	year = {2026},
	pages = {912--917},
}

@article{Shin2021,
  title = {Quantum paraelectric phase of {SrTiO$_3$} from first principles},
  author = {Shin, Dongbin and Latini, Simone and Sch\"afer, Christian and Sato, Shunsuke A. and De Giovannini, Umberto and H\"ubener, Hannes and Rubio, Angel},
  journal = {Phys. Rev. B},
  volume = {104},
  issue = {6},
  pages = {L060103},
  numpages = {6},
  year = {2021},
  month = {Aug},
  publisher = {American Physical Society},
}

@Article{Latini2021,
author = {Simone Latini  and Dongbin Shin  and Shunsuke A. Sato  and Christian Schäfer  and Umberto De Giovannini  and Hannes Hübener  and Angel Rubio },
title = {The ferroelectric photo ground state of {SrTiO$_3$}: Cavity materials engineering},
journal = {Proceedings of the National Academy of Sciences},
volume = {118},
number = {31},
pages = {e2105618118},
year = {2021},
}

@article{Giannozzi2009,
title = "QUANTUM ESPRESSO: A modular and open-source software project for quantum simulations of materials",
author = "Paolo Giannozzi and Stefano Baroni and Nicola Bonini and Matteo Calandra and Roberto Car and Carlo Cavazzoni and Davide Ceresoli and Chiarotti, \{Guido L.\} and Matteo Cococcioni and Ismaila Dabo and \{Dal Corso\}, Andrea and \{De Gironcoli\}, Stefano and Stefano Fabris and Guido Fratesi and Ralph Gebauer and Uwe Gerstmann and Christos Gougoussis and Anton Kokalj and Michele Lazzeri and Layla Martin-Samos and Nicola Marzari and Francesco Mauri and Riccardo Mazzarello and Stefano Paolini and Alfredo Pasquarello and Lorenzo Paulatto and Carlo Sbraccia and Sandro Scandolo and Gabriele Sclauzero and Seitsonen, \{Ari P.\} and Alexander Smogunov and Paolo Umari and Wentzcovitch, \{Renata M.\}",
year = "2009",
doi = "10.1088/0953-8984/21/39/395502",
language = "English (US)",
volume = "21",
journal = "Journal of Physics Condensed Matter",
issn = "0953-8984",
publisher = "Institute of Physics",
number = "39",
}

@article{Perdew1996,
  title = {Generalized Gradient Approximation Made Simple},
  author = {Perdew, John P. and Burke, Kieron and Ernzerhof, Matthias},
  journal = {Phys. Rev. Lett.},
  volume = {77},
  issue = {18},
  pages = {3865--3868},
  numpages = {0},
  year = {1996},
  month = {Oct},
  publisher = {American Physical Society},
}

@article{Svendsen2025,
  title={Effective equilibrium theory of quantum light-matter interaction in cavities for extended systems and the long wavelength approximation},
  author={Svendsen, Mark Kamper and Ruggenthaler, Michael and H{\"u}bener, Hannes and Sch{\"a}fer, Christian and Eckstein, Martin and Rubio, Angel and Latini, Simone},
  journal={Communications Physics},
  volume={8},
  number={1},
  pages={425},
  year={2025},
  publisher={Nature Publishing Group UK London}
}

@article{I-Te2025,
author = {I-Te Lu and Dongbin Shin and Mark Kamper Svendsen and Simone Latini and Hannes H\"{u}bener and Michael Ruggenthaler and Angel Rubio},
journal = {Adv. Opt. Photon.},
number = {2},
pages = {441--525},
publisher = {Optica Publishing Group},
title = {Cavity engineering of solid-state materials without external driving},
volume = {17},
month = {Jun},
year = {2025},
doi = {10.1364/AOP.544138},
}

@article{Fausti2026,
  title={Cavity-enhanced superconducting response in an underdoped cuprate},
  author={Montanaro, Angela and Plastovets, Vadim and Khatiwada, Nitesh and Fiore, Jacopo and Jarc, Giacomo and Alabbadi, Abdullah and Mastropasqua, Antonio and Rigoni, Enrico Maria and Mathengattil, Shahla Y and Zilio, Simone Dal and others},
  journal={arXiv:2606.18084},
  year={2026}
}

@article{Gao2026,
  title={Cavity Enhanced Superconductivity},
  author={Zhang, Hanxiang and Feng, Zexin and Lu, I-Te and Li, Zhiwei and Guo, Songhao and Shang, Qiuyu and Luan, Dening and Panmai, Mingcheng and Watanabe, Kenji and Taniguchi, Takashi and others},
  journal={arXiv:2606.19171},
  year={2026}
}

@article{Junichiro2026,
  title={Vacuum-dressed superconductivity in NbN observed in a high-$ Q $ terahertz cavity},
  author={Xu, Hongjing and Baydin, Andrey and Yi, Qinyan and Lu, I-Te and Zhu, Ningxu and Kritzell, T Elijah and Doumani, Jacques and Kim, Dasom and Tay, Fuyang and Rubio, Angel and others},
  journal={arXiv:2601.08191},
  year={2026}
}

@article{scikit-image,
 title = {scikit-image: image processing in {P}ython},
 author = {van der Walt, {S}t\'efan and {S}ch\"onberger, {J}ohannes {L}. and
           {Nunez-Iglesias}, {J}uan and {B}oulogne, {F}ran\c{c}ois and {W}arner,
           {J}oshua {D}. and {Y}ager, {N}eil and {G}ouillart, {E}mmanuelle and
           {Y}u, {T}ony and the scikit-image contributors},
 year = {2014},
 month = {6},
 volume = {2},
 pages = {e453},
 journal = {PeerJ},
 issn = {2167-8359},
 doi = {10.7717/peerj.453}
}

\newpage
\begin{figure}[h!]
\includegraphics[width=\textwidth]{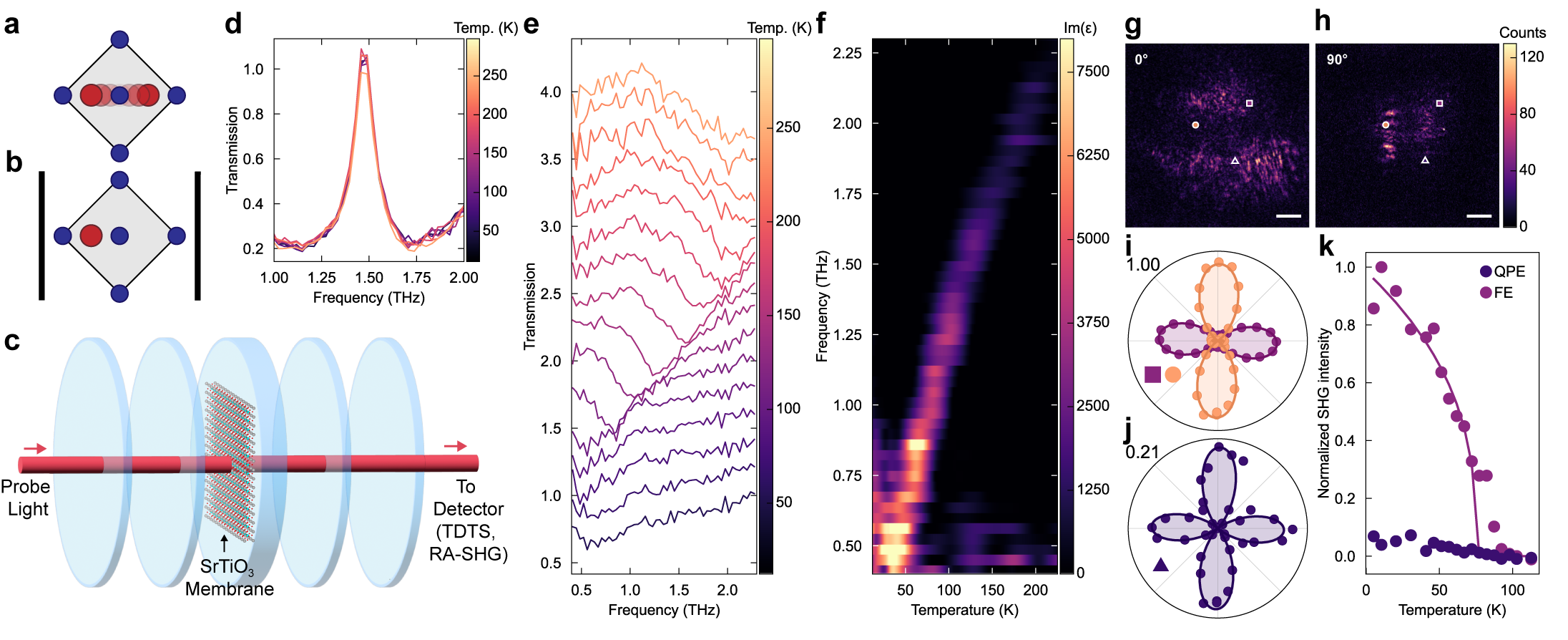}
\label{Fig1}
\end{figure}
\noindent\textbf{Figure 1 $|$ Device architecture and bare PCC and \STO\ properties.} \textbf{a,} Schematic of the QPE phase of \STO. Unlike the ferroelectric phase, where the titanium atom (red) is statically displaced from the center of the oxygen (blue) octahedron, in the QPE phase quantum fluctuations of the FES drive rapid tunneling between the two degenerate displacement configurations---indicated as motion blur---resulting in zero net polarization. \textbf{b,} In the cavity (black lines), coupling to the FES suppresses quantum fluctuations and stabilizes a ferroelectric ground state. \textbf{c,} Schematic of the \STO-PCC device architecture. The PCC consists of diamond layers forming a THz Bragg reflector, with a central defect layer of double thickness that also serves as the substrate for the free-standing \STO\ membrane. The optical transparency of diamond enables both THz and visible-frequency optical probe access. \textbf{d,} Temperature-dependent THz transmission of the PCC. \textbf{e,} Temperature-dependent THz transmission of a $\sim$37 nm thick \STO\ membrane on a silicon substrate. \textbf{f,} Temperature-dependence of $Im(\varepsilon)$ calculated from the data in panel e (Methods). \textbf{g, h,} RA-SHG images at polar angles of 0$^{\circ}$ (g) and 90$^{\circ}$ (h),  corresponding to local maxima of the SHG intensity. Scale bars denote a 10 $\mu$m distance. The data was collected at 5 K. \textbf{i,} SHG polarimetry patterns obtained by integrating the signal over $1\mu\text{m} \times 1\mu\text{m}$ regions (indicated as a circle and square in panels e and g) as a function of polar angle. Magenta and orange correspond to the square and circle regions, respectively. \textbf{j,} SHG polarimetry pattern obtained from the $1\mu\text{m} \times 1\mu\text{m}$ region marked with a dark purple triangle in panels e and g.  \textbf{k,} Temperature-dependent SHG intensity at a polar angle of 0$^{\circ}$, spatially integrated over the same regions as the triangle (QPE, dark purple) and square (FE, magenta) markers in panels i and j. The solid line is a fit to a power law function.

\newpage
\begin{figure}[h!]
\includegraphics[width=89mm]{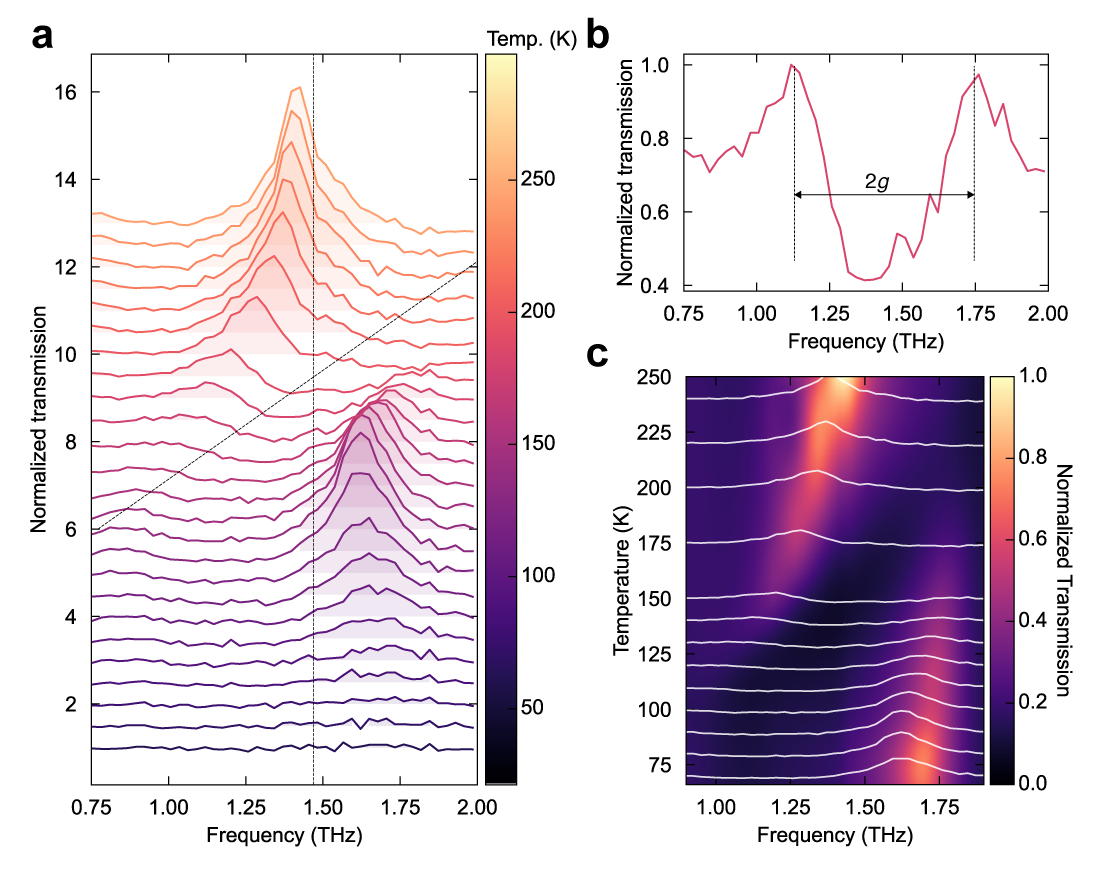}
\label{Fig2}
\end{figure}
\noindent\textbf{Figure 2 $|$ Dielectric response of a 1.5 THz \STO-PCC device.} \textbf{a,} Temperature-dependent THz transmission spectra of the 1.5 THz \STO-PCC device. The vertical and slanted dotted lines are linear fits tracking the bare dispersions of the PCC and \STO\ membrane, respectively. The spectra are normalized by data collected at 15 K (Methods). \textbf{b,} Line cut from data in panel a at 142 K. The dotted lines mark the central frequencies of the LP and UP, which are separated by $2g$ at the avoided crossing. \textbf{c,} Temperature-dependent transfer matrix simulation of THz transmission of the \STO-PCC device (Methods). The simulated data is displayed in the heat map, while the experimental data from panel a is overlaid as white curves. As with the experimental data in panel a, the simulated spectra are normalized by the simulated spectrum at 15 K.\\

\newpage
\begin{figure}[h!]
\includegraphics[width=89mm]{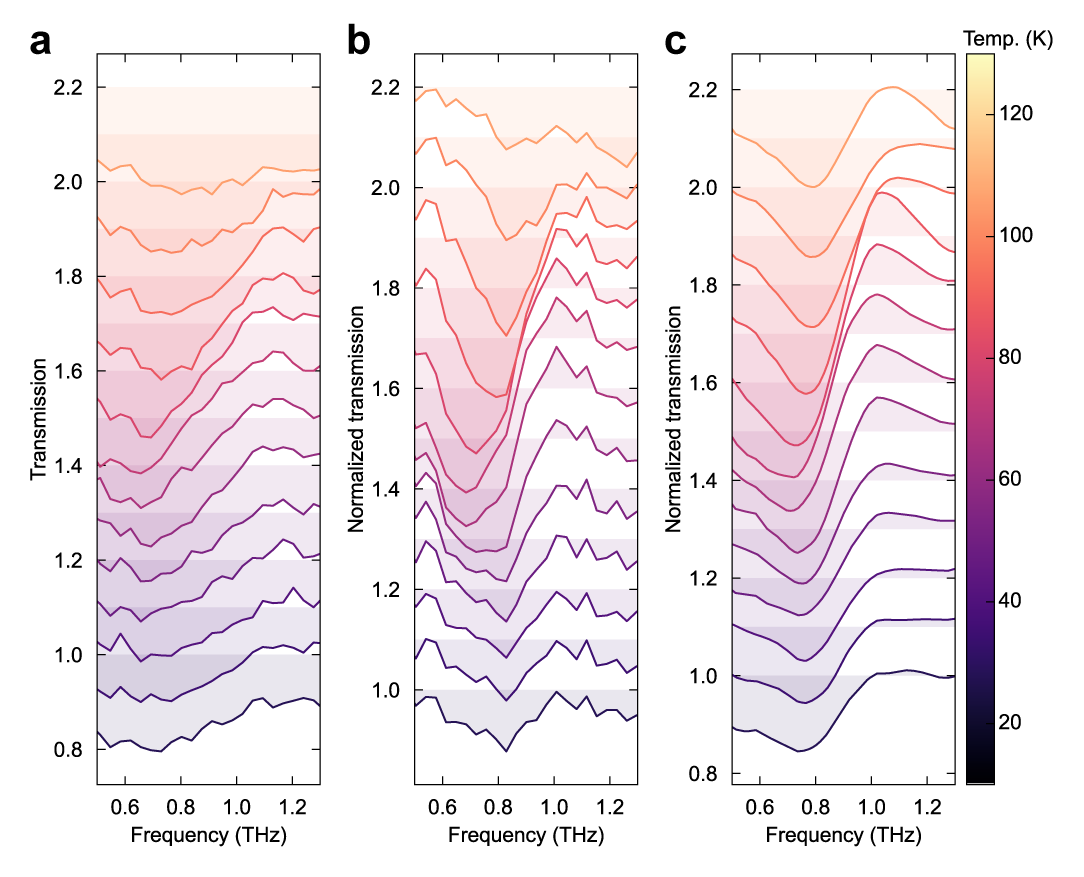}
\label{Fig3}
\end{figure}
\noindent\textbf{Figure 3 $|$ Dielectric response of a 2.5 THz \STO-PCC device.} \textbf{a,} Temperature-dependent THz transmission spectra of the bare 17 nm \STO\ membrane used in this device. \textbf{b,} Temperature-dependent THz transmission spectra of the 2.5 THz \STO-PCC device. The spectra is normalized by the data collected at 300 K (Methods). \textbf{c, } Temperature-dependent transfer matrix simulation of THz transmission of the \STO-PCC device (Methods). As with the experimental data in panel b, the simulated spectra are normalized by the simulated spectrum at 300 K.

\newpage
\begin{figure}[h!]
\includegraphics[width=\textwidth]{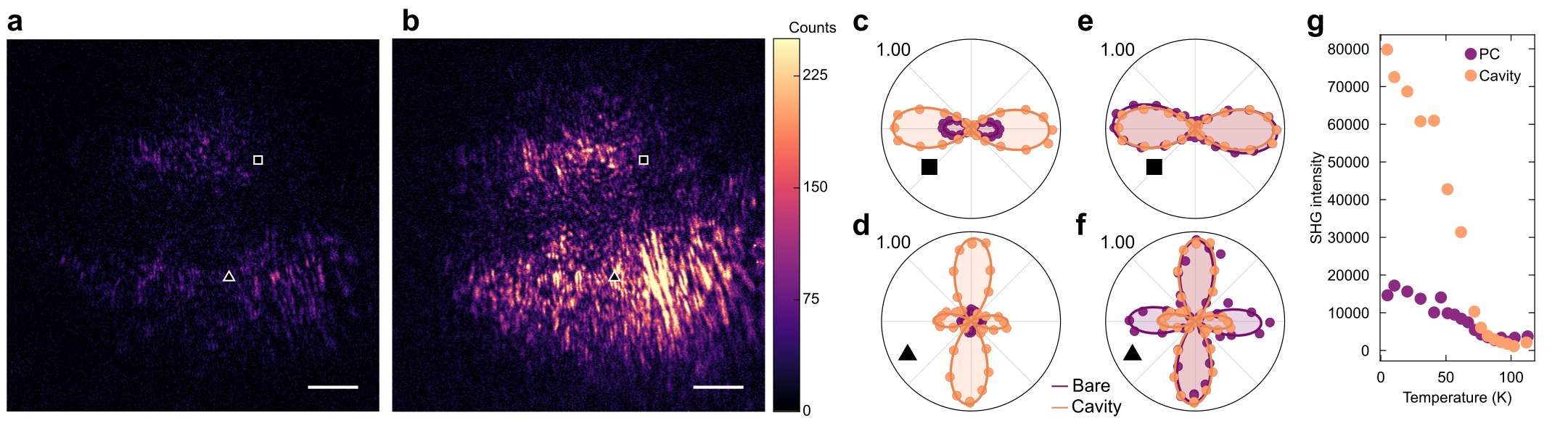}
\label{Fig4}
\end{figure}
\noindent\textbf{Figure 4 $|$ Comparison of the bare and cavity-induced RA-SHG response of a \STO\ membrane.} \textbf{a,} SHG image of a bare 32 nm \STO\ membrane. \textbf{b,} SHG image of a 32 nm \STO\ membrane in cavity-coupled geometry. Both SHG images were collected on the same spot on the sample at 5 K and the scale bars denote a 10 $\mu$m distance. \textbf{c, d, } Comparison of the SHG polarimetry plots in the PC (purple) and PCC (orange) configurations taken from the $1 \mu\text{m} \times 1 \mu\text{m}$ regions marked by a square (c) and triangle (d) in panels a and b. \textbf{e, f, } Same data as in panels c and d, but each individual curve was normalized to its own maximum. These polar plots were collected at 5 K. \textbf{g, } Temperature dependence of the region denoted by a triangle in panels a and b. The data was collected at a polar angle of 90$^{\circ}$. \\

\newpage
\begin{figure}[h!]
\includegraphics[width=\textwidth]{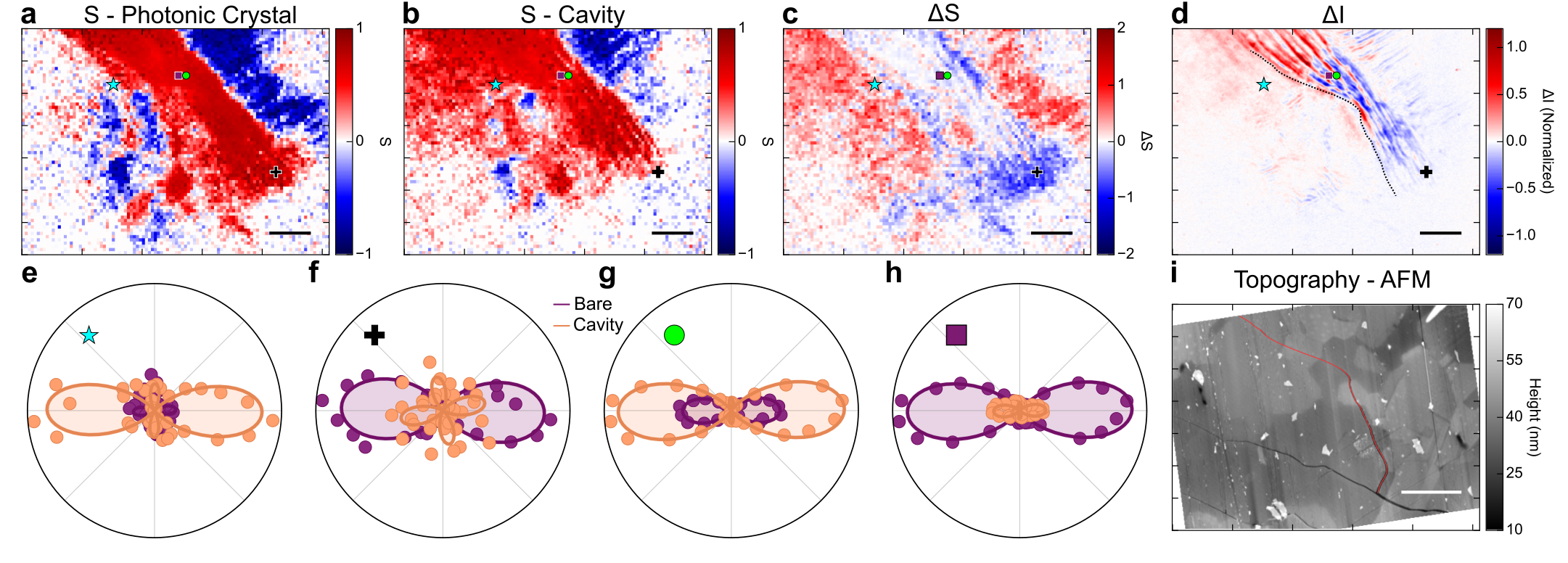}
\label{Fig5}
\end{figure}
\noindent\textbf{Figure 5 $|$ Cavity-modified domain structure of a \STO\ membrane. } \textbf{a,} Spatial map of the symmetry parameter $S$ for a bare \STO\ membrane with a thickness of 18 nm.  \textbf{b, } Spatial map of $S$ for the same region as in panel a, but in the cavity-coupled geometry. \textbf{c, } The change in $S$ obtained by subtracting the data in panel b from that of panel a. \textbf{d, } The normalized angle-summed SHG intensity difference between the cavity-coupled and bare sample for the same region plotted in the other panels. The dotted line tracks the topographic feature highlighted in red in panel i. \textbf{e-h, } Comparison of RA-SHG patterns obtained in the bare (purple) and cavity-coupled (orange) geometries, integrated over the regions designated by the corresponding shapes in panels a-d. \textbf{i, } Topographic AFM map obtained in the same region as panels a-d. The red lines is a guide to the eye that tracks a crack in the sample. The same line is plotted in panel d. All scale bars indicate a distance of 10 $\mu$m and all data was collected at a base temperature of $\sim$5 K.\\

\newpage
\begin{figure}[h!]
    \includegraphics[width=0.85\linewidth]{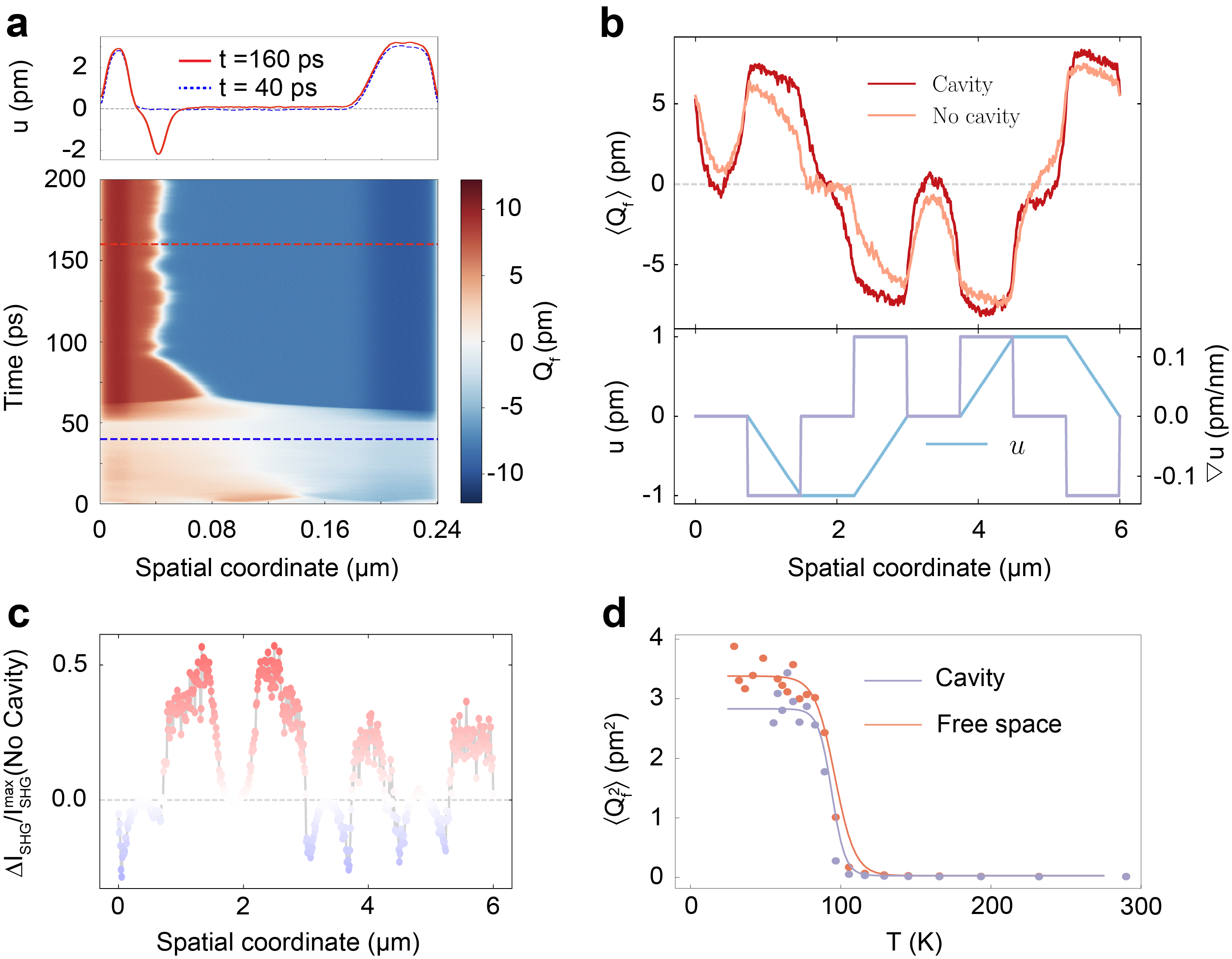}
    \label{Fig6}
\end{figure}
\noindent{\textbf{Figure 6 $|$ Multiscale description of cavity modified ferroelectricity.} \textbf{a,} Time evolution of the local $Q_f$ profile of each cell with the photon field after $50$~ps. The top panels show the corresponding strain distribution. 
    \textbf{b,} Average polarization $\langle Q_f\rangle$ for a two-dimensional system with constant strain along the $y$-direction, and varying strain profile along the $x$-direction. The top panel shows a cut of the data along the line $y = 1.9$ $\mu$m (see SI Figure~S11 for the full 2-D map), while the bottom panel shows the strain profile (blue) and its gradient (purple). 
    \textbf{c,} Difference in second harmonic generation (SHG) intensity inside and outside the cavity, along the same cut as in panel b. 
    \textbf{d,} Average magnitude $\langle Q_f^2\rangle$ as a function of temperature for a sample with uniform strain. The dots are numerical data while the solid curves are fits to the function $a/[1 + e^{b(T - T_c)}] + c$. In all panels, the cavity effect is modeled as an increase of the phonon mass from $M_0$ to $5M_0/4$.}

\newpage

\begin{figure}[H]
    \centering
    \includegraphics[width=0.8\linewidth]{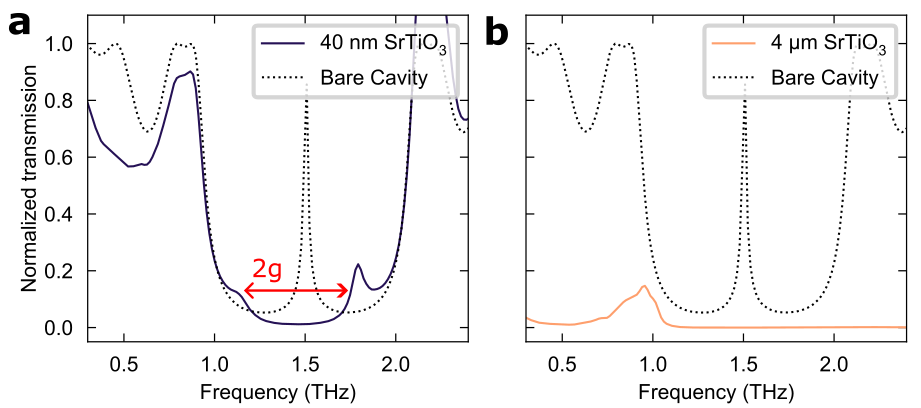}
    \label{fig:thickness_dependence}
\end{figure}
\noindent{\textbf{Extended Data Figure 1 $|$ Dependence of polariton formation on \STO\ thickness.} \textbf{a,b,} A transfer matrix calculation is performed to simulate the spectrum of 40 nm (a) and 4 $\mu$m (b) thick \STO\ samples in a photonic crystal cavity with a central frequency of 1.5 THz. The 40 nm sample depicted in (a) shows the clear splitting associated with polariton formation, while the 4 $\mu$m thick sample in (b) is absent of any clear features, indicating that the thicker \STO\ layer prevents polariton formation. The dielectric function of \STO\ measured at 140 $K$ was used for this simulation.}

\newpage

\begin{figure}[H]
    \centering
    \includegraphics[width=0.9\linewidth]{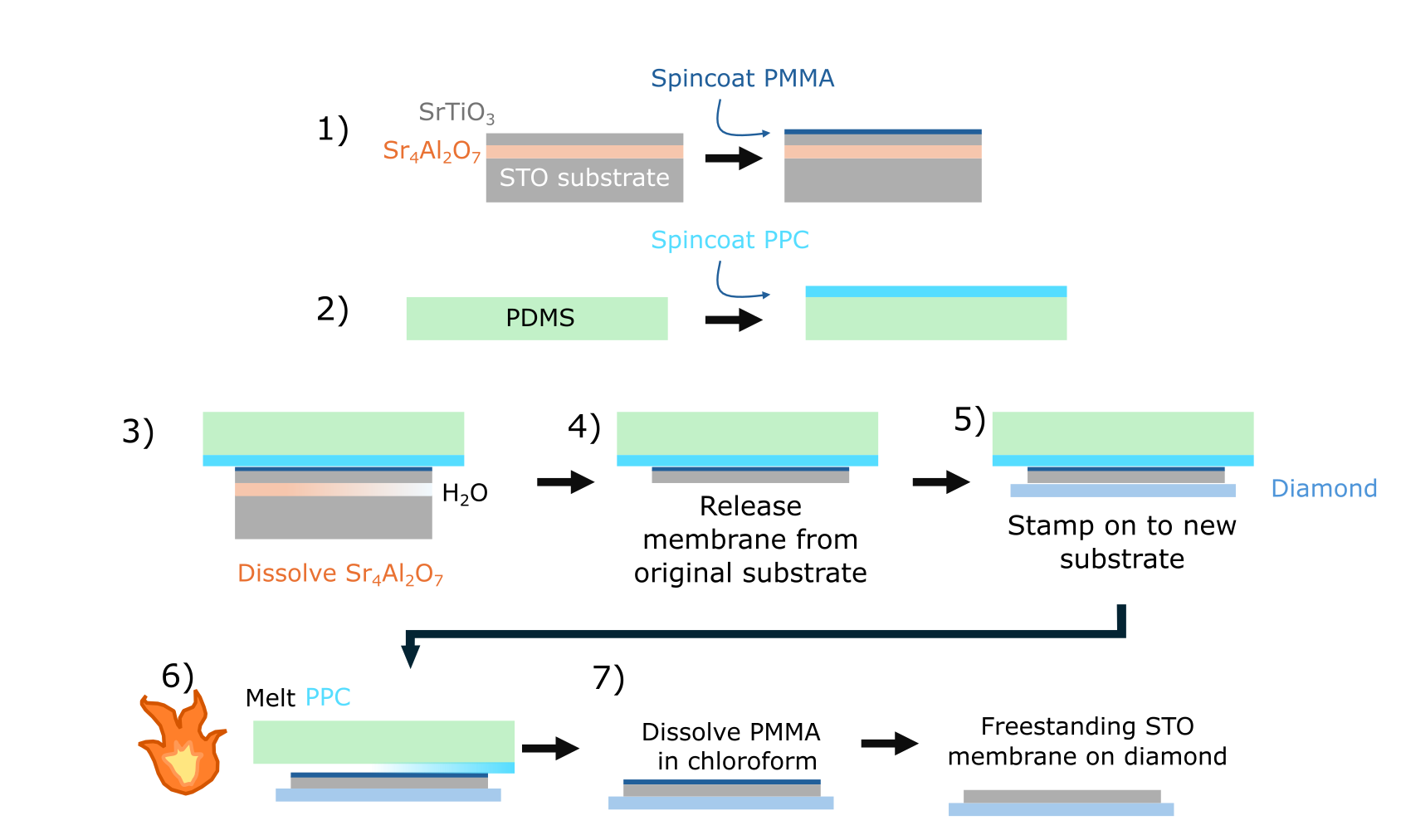}
    \label{fig:membrane_synth}
\end{figure}
\noindent{\textbf{Extended Data Figure 2 $|$ Membrane synthesis procedure.} \STO\ membranes are created by growing \STO\ on a water soluble buffer layer. The transfer of the membrane onto the desired substrate is done via a series of supporting polymers to preserve the structural integrity of the membrane after it is released from its original substrate. The steps are described as follows: \textbf{(1)} the sample has a supporting layer of polymethyl methacrylate (PMMA) spin coated onto the surface to support the sample. \textbf{(2)} A polydimethylsiloxane (PDMS) stamp is prepared by spin coating polypropylene carbonate (PPC) onto its surface and the sample is then put on the stamp before the PPC is fully set. \textbf{(3-4)} The membrane is released on the stamp by dissolving the water soluble buffer layer. \textbf{(5-6)} The membrane is stamped on the desired substrate and released from the stamp by melting the PPC with heat. \textbf{(7)} The surface of the sample is cleaned with chloroform, acetone, and isopropyl alcohol to dissolve residual polymers.}

\newpage 

\begin{figure}[H]
    \centering
    \includegraphics[width=0.9\linewidth]{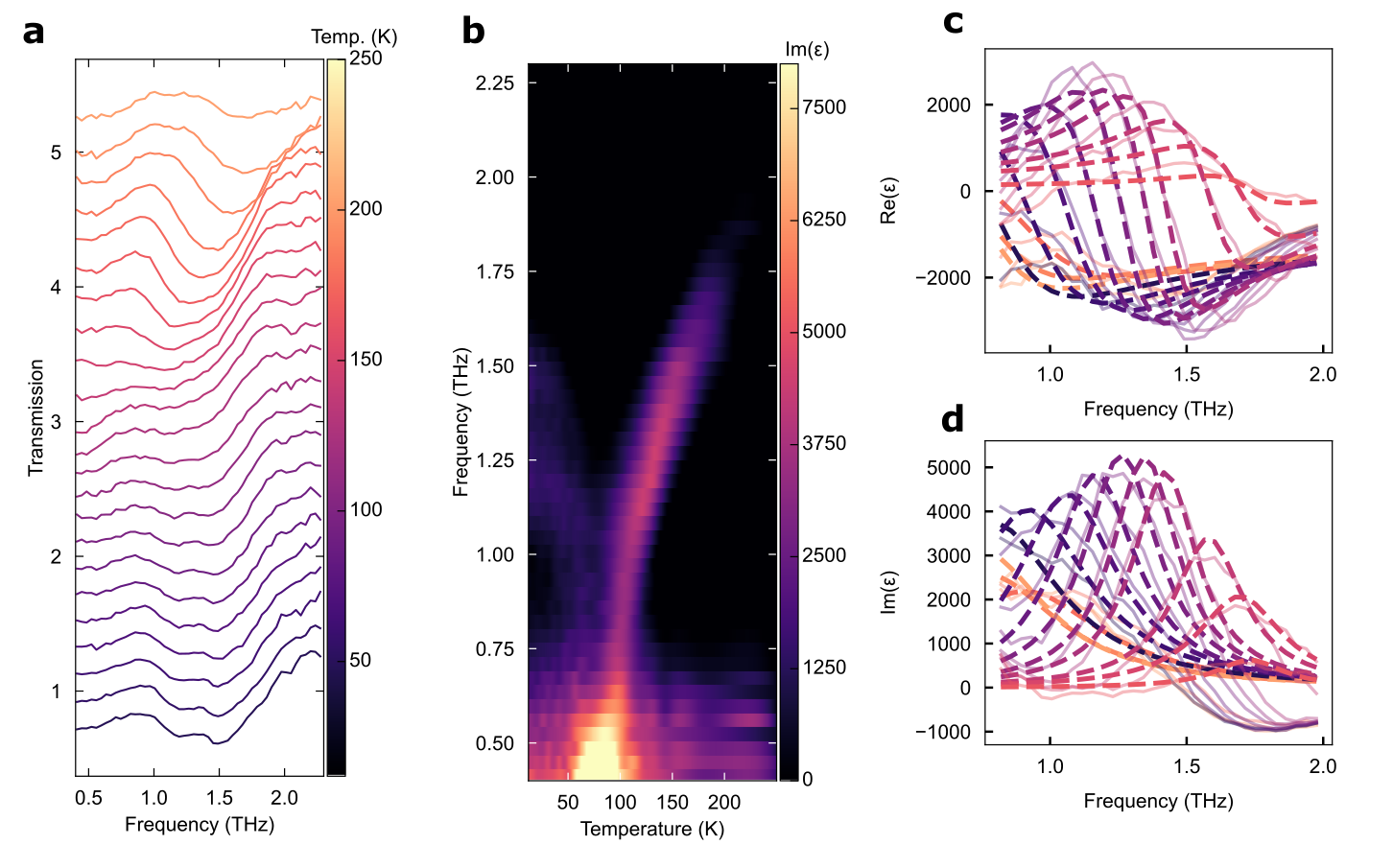}
    \label{fig:bare_sim_params}
\end{figure}
\noindent{\textbf{Extended Data Figure 3 $|$ Bare sample properties for the device depicted in Main Text Figure 2.} \textbf{a,} Temperature-dependent THz transmission spectrum of the bare sample. \textbf{b,} Temperature dependence of $Im(\varepsilon)$ calculated from (a). \textbf{c,d,}  The real (c) and imaginary (d) dielectric functions extracted from (a) at each temperature. The dashed lines are fits to a single Lorentzian model.} 

\newpage 

\begin{figure}[H]
    \centering
    \includegraphics[width=0.4\linewidth]{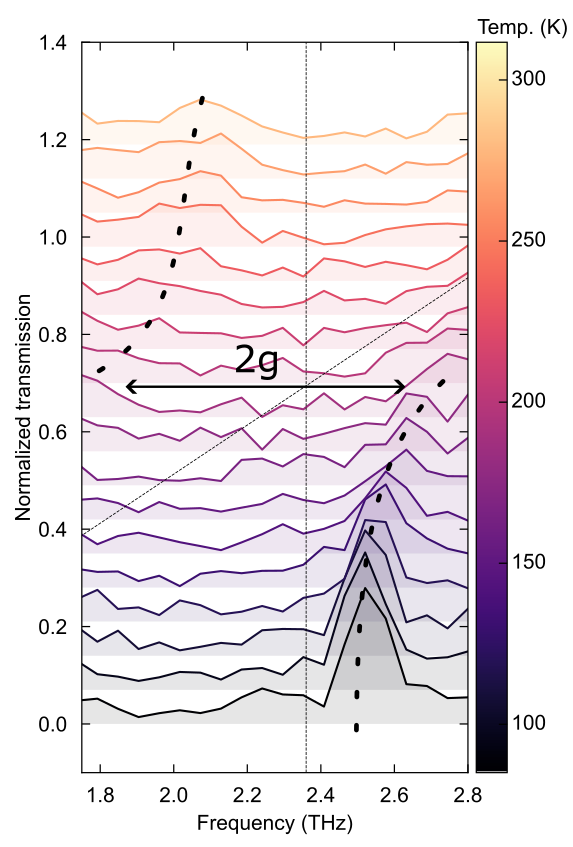}
    \label{fig:other_polariton}
\end{figure}
\noindent{\textbf{Extended Data Figure 4 $|$ Ultrastrong coupling in a second \STO-PCC device.} The temperature dependent transmission spectrum of a second \STO\ cavity device showing ultrastrong coupling ($g/\omega \approx 0.16$). The avoided crossing occurs at 206 K. This device has a central frequency of 2.36 THz and a sample thickness of $\sim$30 $nm$. Thin black dotted lines are guides to the eye depicting the bare cavity frequency and the FES dispersion of the bare sample. The upper and lower polariton are highlighted with thick dotted lines as a guide to the eye. Due to the higher frequency of this cavity there was not an effective way to normalize out the contribution from the band gap of the photonic crystal, so only the data within the band gap is depicted.} 

\newpage

\begin{figure}[H]
    \centering
    \includegraphics[width=0.6\linewidth]{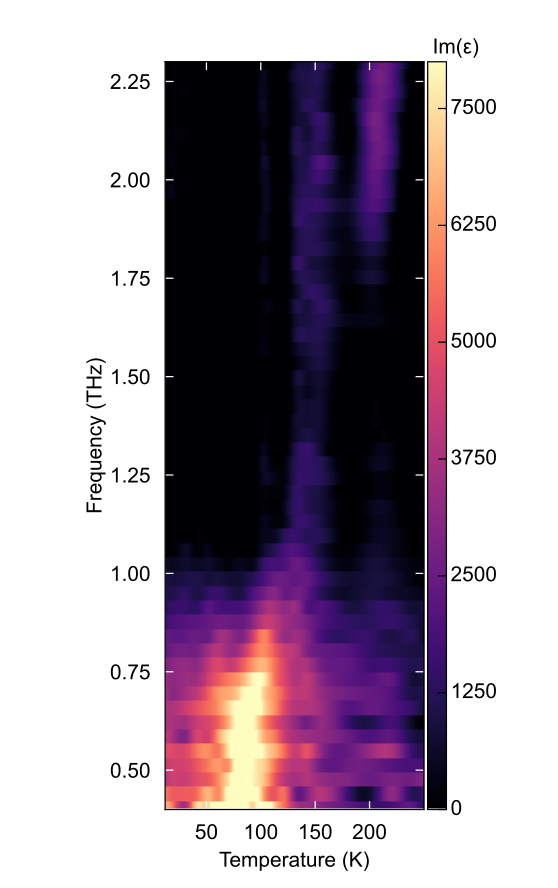}
    \label{fig:dielectric_func}
\end{figure}
\noindent{\textbf{Extended Data Figure 5 $|$ Temperature dependent $Im(\varepsilon)$ spectra of the bare \STO\ used in the device shown in Figures 3, 4, and 5 of the main text.}} 
\newpage

\begin{figure}[H]
    \centering
    \includegraphics[width=0.9\linewidth]{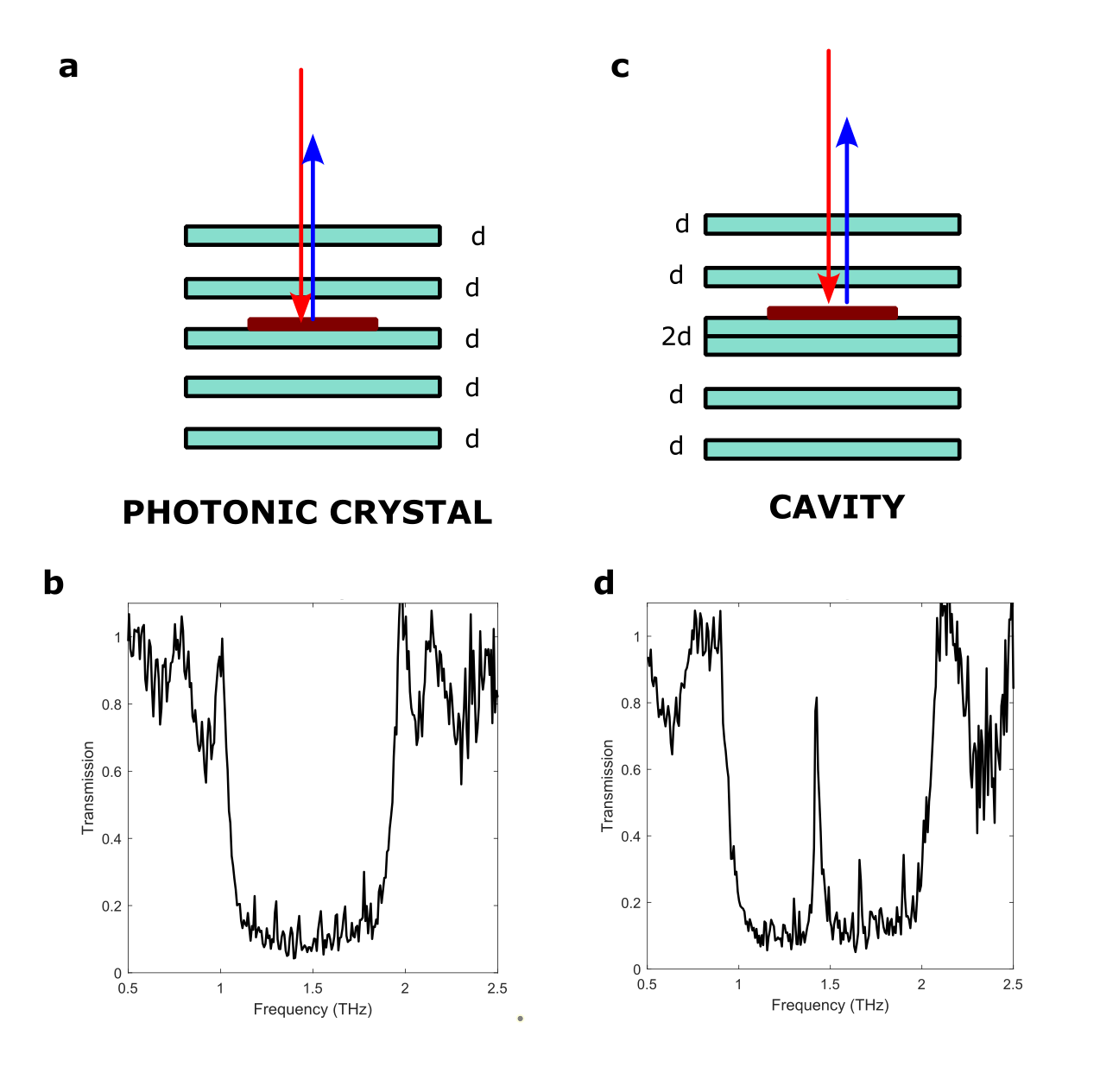}
    \label{fig:geometry}
\end{figure}
\noindent{\textbf{Extended Data Figure 6 $|$ Alternate measurement geometry for RA-SHG imaging measurements.} To control for the objects in the light path during the bare and cavity-coupled measurements, the measurement geometries depicted in panels (a) and (c) were used. \textbf{a, b} In the photonic crystal case, all diamond layers in the structure have the same thickness $d$ (a). As such, no resonance appears in the THz transmision spectrum of the bare photonic crystal, as there is no defect layer (b). \textbf{c, d,} On the other hand, in the cavity configuration, a second diamond layer is added beneath the central layer, creating a total layer thickness of $2d$ (c). Since this structure now has a defect layer, a cavity resonance appears in the bare structure (d). The red squares in (a) and (c) schematically depict where the \STO\ membrane is placed. 

\newpage

\begin{figure}[H]
    \centering
    \includegraphics[width=0.9\linewidth]{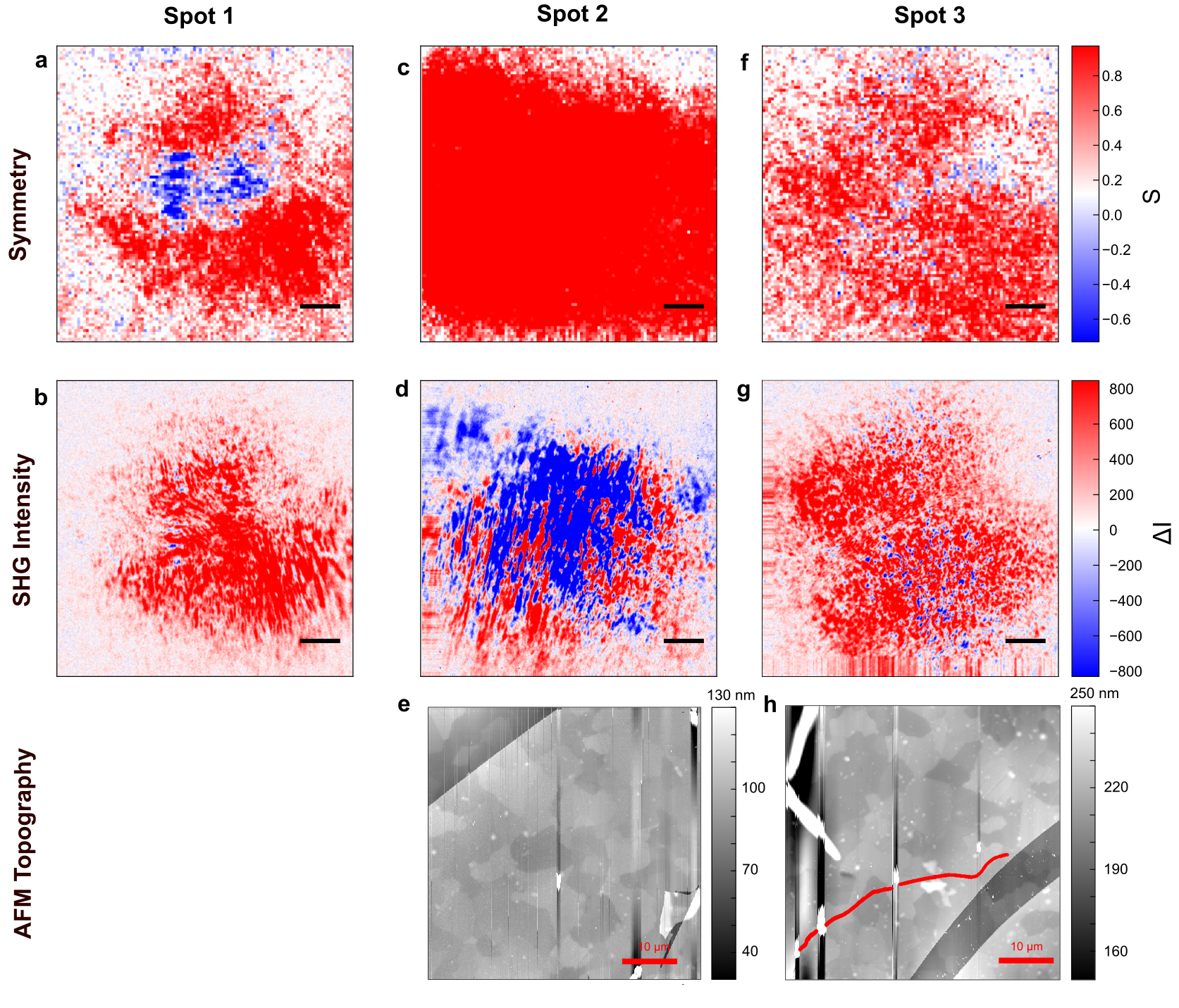}
        \label{fig:other_intensity_var}
\end{figure}
\noindent{\textbf{Extended Data Figure 7 $|$ RA-SHG imaging from multiple devices.} \textbf{a,b,} The spatial distribution of the symmetry parameter $S$ (a) and the intensity change (b) for a spot on the sample depicted in Figure 4 of the main text. \textbf{c,d} The spital distribution of $S$ (c) and the intensity change (d) for a spot (Spot 2) on a different device with a resonance at $\sim$2.5 THz. \textbf{e, } Topographic AFM map of Spot 2. \textbf{f, g, } The spatial distribution $S$ (f) and the intensity change (g) for a second spot (Spot 3) on the same device as Spot 2. \textbf{h,} Topographic AFM map of Spot 3. The red line indicates a crack. All scale bars indicate a distance of 10 $\mu$m and all data was collected at a base temperature of $\sim$5 K. 

\end{document}